\documentclass[12pt]{article}

\usepackage{amsmath,amsfonts,amssymb,mathrsfs,graphicx,cite}

\newcommand{\capdef}{}
\newcommand{\mycaption}[2][\capdef]{\renewcommand{\capdef}{#2}%
        \caption[#1]{{\footnotesize #2}}}
\makeatletter
\renewcommand{\fnum@table}{\textbf{\tablename~\thetable}}
\renewcommand{\fnum@figure}{\textbf{\figurename~\thefigure}}
\makeatother

\newcounter{myenumi}

\renewcommand{\themyenumi}{\roman{myenumi}}
{\end{list}}

\newlength{\myem}
\newcounter{mysubequation}[equation]

\makeatletter
\renewcommand{\section}{\@startsection{section}{1}{0em}{-\baselineskip}%
{\baselineskip}{\normalfont\large\bfseries}}
\renewcommand{\subsection}%
{\@startsection{subsection}{2}{0em}{-0.7\baselineskip}%
{0.7\baselineskip}{\normalfont\bfseries}}
\makeatother

\newcommand{\bi}{\begin{itemize}}
\newcommand{\ei}{\end{itemize}}

\newcommand{\be}{\begin{equation}}
\newcommand{\ee}{\end{equation}}
\newcommand{\bea}{\begin{eqnarray}}
\newcommand{\eea}{\end{eqnarray}}

\newcommand{\ldm}{\Delta m_{31}^2}
\newcommand{\sdm}{\Delta m_{21}^2}
\newcommand{\deltacp}{\delta_{\mathrm{CP}}}
\newcommand{\stheta}{\sin^2  2 \theta_{13} }

\newcommand{\ie}{{\it i.e.}}

\newcommand{\eg}{{\it e.g.}}

\newcommand{\cf}{{\it cf.}}

\newcommand{\etc}{{\it etc.}}
\newcommand{\eq}{Eq.}

\newcommand{\fig}{Fig.}

\newcommand{\Ref}{Ref.}
\newcommand{\Refs}{Refs.}
\newcommand{\Sec}{Sec.}

\newcommand{\App}{Appendix}

\newcommand{\Tab}{Table}

\newcommand{\equ}[1]{\eq~(\ref{equ:#1})}
\newcommand{\figu}[1]{\fig~\ref{fig:#1}}

\begin{document}


\begin{titlepage}

\renewcommand{\thefootnote}{\alph{footnote}}

\vspace*{-3.cm}
\begin{flushright}
\end{flushright}


\renewcommand{\thefootnote}{\fnsymbol{footnote}}
\setcounter{footnote}{-1}

{\begin{center}
{\large\bf How astrophysical neutrino sources could be used for early measurements 
of neutrino mass hierarchy and leptonic CP phase
} \end{center}}
\renewcommand{\thefootnote}{\alph{footnote}}

\vspace*{.8cm}
\vspace*{.3cm}
{\begin{center} {\large{\sc Walter~Winter\footnote[1]{\makebox[1.cm]{Email:}
                winter@ias.edu}
		}}
\end{center}}
\vspace*{0cm}
{\it
\begin{center}

       School of Natural Sciences, Institute for Advanced Study,
       Princeton, NJ 08540

\vspace*{1cm}


\end{center}}

\vspace*{1.5cm}

{\Large \bf
\begin{center} Abstract \end{center}  }

We discuss the possible impact of astrophysical neutrino flux measurements
at neutrino telescopes on the neutrino oscillation program of reactor experiments
and neutrino beams. We consider neutrino fluxes from
neutron sources, muon damped sources, and pion sources, where we parameterize the
input from these sources in terms of the flux ratio $R=\phi_\mu/(\phi_e+\phi_\tau)$
which can be extracted from the muon track to shower ratio in a neutrino telescope.
While it is difficult to obtain any information from this ratio alone, we
demonstrate that the dependence on the oscillation parameters is very 
complementary to the one of reactor experiments and neutrino beams.
We find that for large values of $\sin^2 2 \theta_{13}$, a measurement of $R$ with a precision of 
about 20\% or better may not only improve the measurement of the leptonic CP phase, 
but also help the determination of the mass hierarchy.
In some cases, early information on $\delta_{\mathrm{CP}}$ may even be obtained from
Double Chooz and an astrophysical flux alone without the help of superbeams. 
For small values of $\sin^2 2 \theta_{13}$, we find that using the information from 
an astrophysical neutrino flux could eliminate the octant degeneracy better than reactor experiments
and beams alone. Finally, we demonstrate that implementing an additional observable based on the electromagnetic to hadronic shower ratio at a neutrino telescope (such as at higher
 energies) could be especially beneficial for pion beam sources.

\vspace*{.5cm}

\end{titlepage}

\newpage

\renewcommand{\thefootnote}{\arabic{footnote}}
\setcounter{footnote}{0}


\section{Introduction}

In the coming ten years, many 
experiments with a terrestrial neutrino source will provide information on the neutrino
oscillation parameters. In particular, the parameters $\stheta$, $\deltacp$, and the neutrino mass
hierarchy are very interesting, since we know very little about them
except from an upper bound on $\stheta$~\cite{Apollonio:1999ae}. However, even if $\stheta$
is large, measuring leptonic CP violation and the neutrino mass hierarchy with long-baseline and
reactor experiments will be extremely challenging on that time scale because of the
low statistics (see, \eg, \Ref~\cite{Huber:2004ug}). Therefore, any complementary
information could help to improve the measurement. In this study, we discuss 
the potential complementarity of three sources of information: Neutrino beams, reactor experiments, and a possible flavor composition measurement of an astrophysical neutrino flux.

\subsubsection*{Considered sources}

Neutrino beams, such as MINOS~\cite{Ables:1995wq}, T2K~\cite{Itow:2001ee}, or NO$\nu$A~\cite{Ayres:2004js}, are sensitive to $\stheta$, $\deltacp$,
and the neutrino mass hierarchy via electron neutrino appearance. However, this ability to access
all of the parameters leads to correlations and degeneracies~\cite{Fogli:1996pv,Burguet-Castell:2001ez,Minakata:2001qm,Barger:2001yr} when it comes to the extraction of the individual parameters. In addition, the next generation of ``narrow-band beams'' operated at the oscillation maximum will be mainly sensitive to the CP-odd part of the appearance probability. Reactor experiments~\cite{Anderson:2004pk}, such as Double Chooz~\cite{Ardellier:2004ui}, will be very complementary to provide ``clean'' information on $\stheta$ only~\cite{Minakata:2002jv,Huber:2003pm}. However, these experiments cannot
access $\deltacp$ and the mass hierarchy at all, and the statistics of the first generation of experiments will be moderate.

As another potential source of information,
a detection of astrophysical neutrinos at neutrino telescopes~\cite{Aslanides:1999vq,Ahrens:2002dv,Tzamarias:2003wd,Piattelli:2005hz} with a well-predicted flavor composition at the source (such as from neutron or pion decays) could provide additional knowledge on the mixing parameters~\cite{Serpico:2005sz,Serpico:2005bs,Bhattacharjee:2005nh}.
The existence of such astrophysical neutrinos is not yet proven, but the detection of very high energy
cosmic rays suggests that cosmic accelerators exist which should also produce high energy neutrinos,
such as by pion decays. There are many potential candidates for neutrino sources, such as gamma ray bursts, active galactic nuclei, or starburst galaxies, where the latter source is not affected by the Waxmann-Bahcall bound~\cite{Loeb:2006tw}. Since we are not interesting in the type of source, but the
initial flavor composition in this study, we will follow a more abstract approach and classify potential sources by their initial flavor composition below. If the distance to
the source is long enough or the production region is large enough, neutrino oscillations
on the way from the source to the Earth will average out. The predicted flavor composition
at the Earth then depends on the mixing parameters including $\deltacp$ and the CP-even part of the mixing only (see, \eg, \Ref~\cite{Mena:2003ug}), whereas it is not sensitive to the mass hierarchy. However, compared to the terrestrial neutrino sources, there will be no spectral information, which means that, given the low statistics, it will be difficult to extract some information
on the individual mixing angles. 

For the astrophysical sources, we consider neutrino production from neutron decays with the flavor ratio (1:0:0) at the source~\cite{Anchordoqui:2003vc,Hooper:2004xr} (``Neutron beam source''), from pion decays with the flavor ratio (1:2:0) at the source (``Pion beam source''), and from muon damped pion decays with a flavor ratio (0:1:0) at the source~\cite{Rachen:1998fd,Kashti:2005qa} (``Muon damped source''). In the last case, the muons are absorbed before they can decay, which may also occur for pion beams at high energies~\cite{Kashti:2005qa}. 
The neutrino fluxes expected from these sources typically scale as $dN/dE \propto E^{-2}$, where the
typical upper neutrino energies can be very different.\footnote{Note that since the cross sections increase with energy, the actual event rates are determined by a balance between flux and cross sections.} For example, for gamma ray bursts, the typical
neutrino energies  depend very much on the production region (and process) and can be between GeV and EeV.
For a neutron beam produced by photo-dissociation of heavy nuclei, the typical energies are more limited to the TeV range. The sources can be either steady (such as for a neutron beam produced by photo-dissociation of heavy nuclei, a pion beam from active galactic nuclei, or a diffuse background flux), or time-dependent on a short timescale (such as for a flux from a gamma ray burst). 
Depending on neutrino energies and steadiness of the source, the background especially from atmospheric neutrinos can be problematic for the event extraction especially below about $1 \, \mathrm{PeV}$. 
For more discussion on background discrimination approaches, see \App~\ref{app:stat}.

\subsubsection*{Observables and statistics for astrophysical neutrinos}

As far as the detection of the astrophysical neutrinos is concerned, we do not distinguish between neutrinos and antineutrinos, \ie, the neutrino and antineutrino fluxes are simply added. A neutrino telescope 
can identify muons by their tracks (Cherenkov light with continuous loss of energy).
However, electron and tau neutrino events are harder to disentangle because they both produce showers of particles with a larger threshold. Depending on the neutrino energy, the electromagnetic showers (from electrons) and hadronic showers (from taus) may be disentangled by their muon content, as well as the tau track may be measured, which means that there could be a possibility to disentangle electron and tau neutrino neutrino events (see, \eg, \Refs~\cite{Learned:1994wg,Beacom:2003nh}). An additional process to identify $\bar{\nu}_e$'s is the Glashow resonant process $\bar{\nu}_e + e^- \rightarrow W^-$ at 6.3~PeV, which can also be used for oscillation parameter measurements~\cite{Bhattacharjee:2005nh}.
For most of this study,
we follow \Refs~\cite{Serpico:2005sz,Serpico:2005bs} (see also \Ref~\cite{Dutta:2000jv}) and use $R=\phi_\mu/(\phi_e+\phi_\tau)$
as observable, which means that we make the conservative assumption
that electron and tau events cannot be disentangled. This observable can be extracted
from the ratio of muon tracks to showers~\cite{Beacom:2003nh}, but note
 that there is an additional hadronic shower background from neutral current events for all flavors which needs to be subtracted. We assume that we have a precision measurement of $R$ 
 with an effective relative error exploiting all available information and containing all
 systematics and backgrounds. For example, the information from different energies and event types may be
 used to reduce the systematical errors on that quantity (see, \eg, \Refs~\cite{Anchordoqui:2004eb,Kashti:2005qa}).  Especially using different energies
 is plausible for neutrino telescopes, since we assume that
 neutrino oscillations are averaged out (compared to the beams and reactor experiments, there will be no energy dependence of the oscillations). Finally, we assume a signal without any ``new physics'' contamination, such as neutrino decays~\cite{Beacom:2002vi,Beacom:2003zg,Barenboim:2003jm}.  
 
 For this study, we will use this effective error on $R$ to formulate the requirements to
 a neutrino telescope to be useful for terrestrial\footnote{Further on, we refer to the
 ``terrestrial experiments'' as the experiments with a terrestrial, \ie, Earth-based, neutrino source.} neutrino oscillation experiments, \ie,
 we will show the results for different errors and read off the required precision. 
  Of course, the effective error will not only depend on systematics, but also on the number of events expected. This signal  height depends on the distance and energy release of the event, which means that
  it is not predictable until measured. Typically, several tens of events are expected for
astrophysical neutrino sources, whereas of the order of hundred events can be
expected for electron neutrino appearance at superbeams for large $\stheta \sim 0.1$ (see, \eg, \Ref~\cite{Huber:2004ug}). 
This means that the actual statistical errors from astrophysical sources and
superbeams ($\propto \sqrt{N}$) may not be different by orders of magnitude, 
and a combination of data could make sense -- a hypothesis to be tested in this work. 
We will  typically use $5\%$, $10\%$, and $20\%$ errors on $R$ in order to discuss the 
requirements to an astrophysical measurement in the following sections. For example, a precision
 of $\sim 20\%$ was found in \Ref~\cite{Beacom:2003nh} for a flux close to the Waxman-Bahcall bound~\cite{Waxman:1998yy,Bahcall:1999yr} measured by IceCube. This flux of $E_{\nu_\mu}^2 dN_{\nu_\mu}/dE_{\nu_\mu} = 10^{-7} \, \mathrm{GeV} \, \mathrm{cm}^2 \, \mathrm{s}^{-1}$ was assumed for one
 year at IceCube, which means that higher precisions may be expected for longer running times.
 Larger fluxes and therefore smaller errors on $R$ could also be obtained for IceCube volume upgrades,  optically thick sources, galactic sources, or sources not emitting protons at energies
the Waxman-Bahcall bound is normalized to (see, \eg, \Refs~\cite{Mannheim:1998wp,Loeb:2006tw}).
We illustrate in \App~\ref{app:stat} how to relate such errors on $R$ to specific event rates
in a neutrino telescope, where we also discuss principle challenges for backgrounds, systematics, and the different sources.

\subsubsection*{Astrophysical source identification}

We assume that the type of the astrophysical source can be identified. 
The results of this study are then to be interpreted together with the assumption of that source. 
In order to identify the source, first of all note that the ratios $R$ for the discussed sources are strongly separated even for different oscillation parameters, which means that a small error on $R$ would instantly be a strong hint for the source determination
($R \sim 0.26$, $\sim 0.66$, and $\sim 0.5$, respectively, for neutron beams, muon damped sources, and pion beams, as well as $\stheta=0$). In addition, the energy dependence can be used for the source
determination because the flavor composition may change as function of energy in a characteristic, source-dependent way, such as from a pion beam to a muon damped source~\cite{Kashti:2005qa}. 
Furthermore, associated signals in gamma or cosmic rays can be used for source identification (see, \eg, \Ref~\cite{Serpico:2005sz}). Different flavor ratios also help to identify the source (see, \eg, \Ref~\cite{Xing:2006uk} even for arbitrary flavor compositions), and the Glashow resonance mentioned above may help as well. If an independent source identification is missing,
only generic methods, such as in \Ref~\cite{Serpico:2005bs}, can be used to infer on the neutrino oscillation parameters.
In addition, we assume that there is little contamination from different production mechanisms producing different flavor ratios at the source. For example, neutrons can be generated by collisions of high energy protons on ambient photons and protons, or in the photo-dissociation of heavy nuclei. In the first case, the $\bar{\nu}_e$ from neutron decays would be negligible compared to the neutrino flux from pion decays, which means that we effectively have a pion beam source. In the second case, photo-dissociation of heavy nuclei produces a pure $\bar{\nu}_e$ flux, \ie, we have a neutron beam source. In either case, the obtained flavor composition at the source is relatively clean.
 
In summary, there might be three interesting sources of
information for the neutrino oscillation parameter measurements, which are very complementary,
but all suffer from low precisions on a timescale of the coming ten years: Neutrino beams, reactor experiments, and astrophysical neutrino flux measurements. In this study,
we demonstrate that the combination of the complementary knowledge could allow for
early information on the unknown neutrino oscillation parameters, while the knowledge
and statistics from each individual source will be moderate or poor. Note that, in principle, other
complementary information could be obtained from $0\nu\beta\beta$-decay (for the mass hierarchy, see \eg\ \Ref~\cite{Pascoli:2005zb} and references therein), a galactic supernova explosion (see, \eg, \Refs~\cite{Lunardini:2001pb,Bandyopadhyay:2003ts,Dighe:2003be,Barger:2005it}),
or cosmology (see also \Ref~\cite{Lindner:2005kr} for a different combination of
information). We will not discuss these sources at this place.

This study is organized as follows: First, we illustrate and motivate the complementarity among
superbeams, reactor experiments, and astrophysical sources in \Sec~\ref{sec:compl}.
Then we describe the simulation methods and assumptions used in \Sec~\ref{sec:sim}.
The results sections (\Sec~\ref{sec:reactor} to \Sec~\ref{sec:flavors}) are ordered by
the timescale of relevance and level of technicality: In \Sec~\ref{sec:reactor} we
discuss possible early information on $\deltacp$ using reactor experiments and
astrophysical sources only. In \Sec~\ref{sec:mh}, we illustrate the impact on
the mass hierarchy measurements at superbeams and reactor experiments. Then 
in \Sec~\ref{sec:cp}, we investigate CP violation and CP precision measurements.
Furthermore, we demonstrate in \Sec~\ref{sec:theta23} how and where in parameter space 
the potential to resolve the octant degeneracy could be improved. Finally, 
we discuss the impact of the measurement of all flavors in \Sec~\ref{sec:flavors},
which turns out to be very useful for pion beams. 
 Note that we do not consider precision measurements of $\theta_{12}$ by astrophysical fluxes,
 which have been discussed elsewhere~\cite{Bhattacharjee:2005nh}.
 
\section{Complementarity terrestrial-astrophysical}
\label{sec:compl}

In this section, we illustrate the complementarity between astrophysical sources and
terrestrial neutrino oscillation experiments, \ie, neutrino beams and reactor experiments. 
Unless noted otherwise, we use the following set of simulated neutrino oscillation
parameters throughout this study~\cite{Fogli:2005cq,Bahcall:2004ut,Bandyopadhyay:2004da,Maltoni:2004ei}:
$\ldm = 2.5 \, \cdot 10^{-3} \, \mathrm{eV}^2$, $\sin^2 2
\theta_{23}=1$, $\sdm = 8.2 \cdot 10^{-5} \, \mathrm{eV}^2$, $\sin^2 2
\theta_{12}=0.83$, and $\stheta=0.1$ somewhat below the CHOOZ
bound~\cite{Apollonio:2002gd}, $\deltacp=0$,
and a normal hierarchy.

In order to study $R \equiv \phi_\mu/(\phi_e+\phi_\tau)$ 
for astrophysical
sources, we assume that
either the production or detection of the neutrinos is incoherent,
or the mass eigenstates loose coherence while traveling because of the
long distances (see, \eg, \Ref~\cite{Giunti:1997wq} for a formal treatment). 
Therefore, any oscillations and CP-violating effects 
average out in order to obtain 
\begin{eqnarray}
P_{\alpha \beta}  & = & \sum\limits_{i=1}^3 |U_{\alpha i}|^2 |U_{\beta i}|^2 \nonumber \\
& = & \delta_{\alpha \beta} - 2 \sum\limits_{i<j} \mathrm{Re}(U_{\alpha i} U_{\alpha j}^* U_{\beta i}^* U_{\beta j}) \, .
\label{equ:p}
\end{eqnarray}
In the simplest cases, one has then for $R$
\begin{eqnarray}
R^{\mathrm{Neutron \, beam}} & = & \frac{P_{e \mu}}{P_{ee} + P_{e \tau}} =  \frac{P_{e \mu}}{1 - P_{e \mu}} \, , \nonumber \\
R^{\mathrm{Muon \, damped}} & = & \frac{P_{\mu \mu}}{P_{\mu e} + P_{\mu \tau}} =  \frac{P_{\mu \mu}}{1 - P_{\mu \mu}} \, ,
\end{eqnarray}
whereas for the pion beam
\begin{eqnarray}
R^{\mathrm{Pion \, beam}} & = & \frac{2 P_{\mu \mu}+P_{e \mu}}{2 P_{\mu e} + P_{ee} + 2 P_{\mu \tau}+P_{e \tau}} \, .
\end{eqnarray}
Note that the probabilities in these formulas are the averaged ones from \equ{p}.
Using the standard parameterization of the mixing matrix $U$~\cite{Eidelman:2004wy} 
and our standard values of the other oscillation parameters, one can then calculate $R$
as function of the oscillation parameters for the different sources. We expand $R$ for the
different astrophysical sources to first order in $\theta_{13}$:
\begin{eqnarray}
R^{\mathrm{Neutron \, beam}} & \sim & 0.26 + 0.30 \, \,  \theta_{13} \, \cos \deltacp \, , \\
R^{\mathrm{Muon \, damped}} & \sim & 0.66 - 0.52 \, \,  \theta_{13} \, \cos \deltacp  \, , \\
R^{\mathrm{Pion \, beam}} & \sim & 0.50 - 0.14 \, \,  \theta_{13} \, \cos \deltacp \, . 
\end{eqnarray}
Higher order terms in $\theta_{13}$ are relatively small (but not negligible). For neutrino beams, however, the $\deltacp$-dependent terms in $P_{\mu e}$ 
are suppressed by the mass hierarchy, which means that the $\theta_{13}^2$-term is 
the leading term for large $\theta_{13}$. At the first oscillation maximum and in vacuum, we
find (see, \eg, \Ref~\cite{Akhmedov:2004ny}):
\begin{equation}
P_{\mu e} \sim 2 \, \theta_{13}^2 \pm  0.09 \, \, \theta_{13} \sin \deltacp \, ,
\end{equation}
where the plus is for antineutrinos and the minus for neutrinos.
Because most of the first-generation superbeams are operated close to the
first oscillation maximum and have a very narrow beam spectrum, this approximation
should be useful for qualitative discussions. Most importantly, neutrino beams 
are dependent on the CP-odd $\sin \deltacp$, whereas astrophysical sources 
are dependent on the CP-even $\cos \deltacp $. In addition, the amplitude of the
signal for neutrino beams will be determined by $\theta_{13}^2$, whereas 
$\theta_{13}$ acts as a perturbation for astrophysical sources. 
The astrophysical information does not depend on neutrinos or antineutrinos,
whereas neutrino beams usually exploit exactly this dependence by implementing 
neutrino and antineutrino channels. Finally, pion and
muon-damped sources have a different sign in front of the $\cos \deltacp$-term than
neutron beam sources, and therefore the roles of $\deltacp=0$ and $\pi$ are exchanged.

\begin{figure}[t!]
\begin{center}
\includegraphics[width=0.9\textwidth]{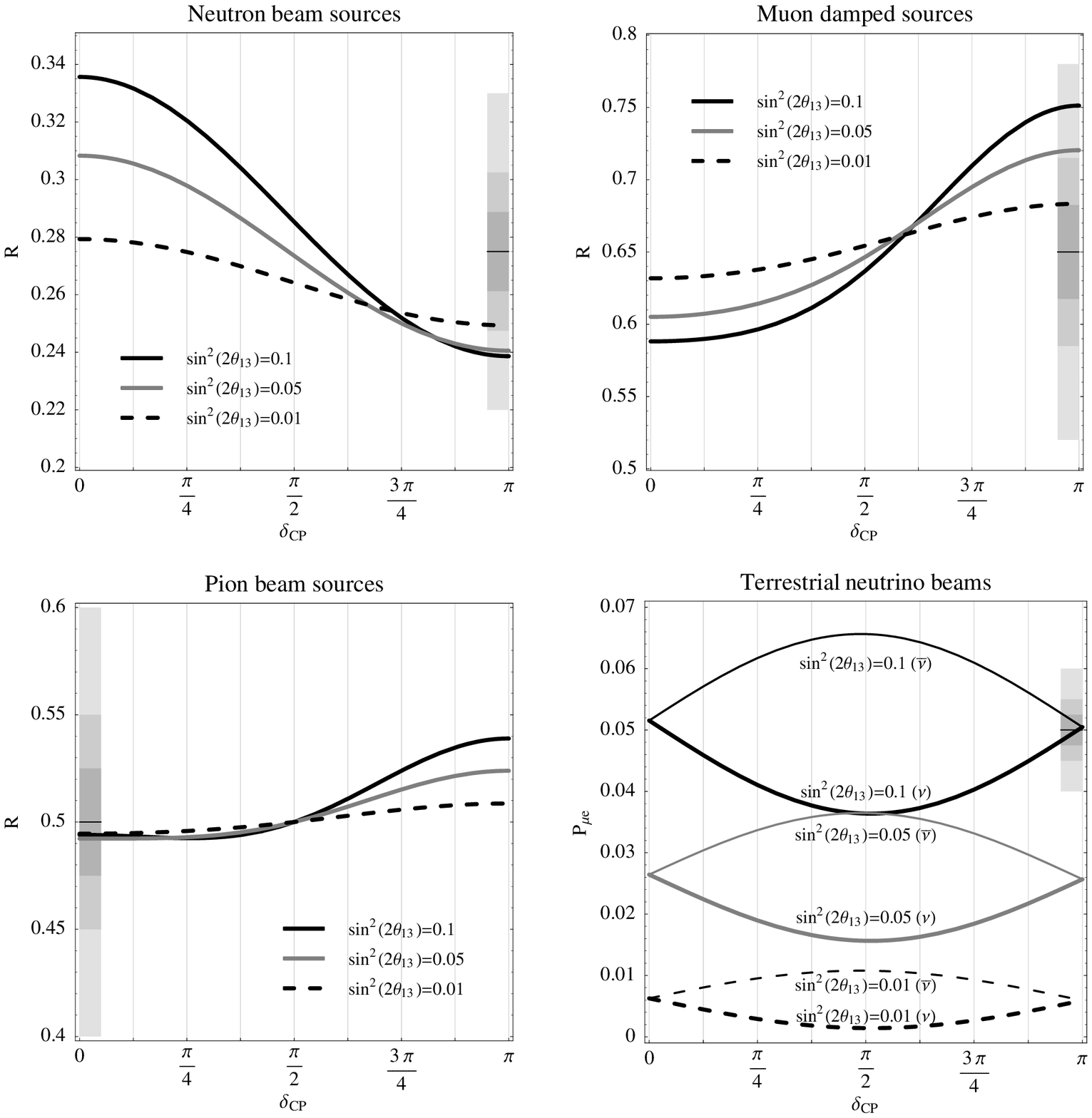}
\end{center}
\mycaption{\label{fig:sources} Sources used in this study for which the signal depends on $\deltacp$.
We show the quantities $R$ and $P_{\mu e}$ ($P_{\bar{\mu} \bar{e}}$), respectively, as function of
 $\deltacp$ for different values of $\stheta$. The shaded bars illustrate the size of the $5\%$, $10\%$, and $20\%$ errors for the chosen central values (horizontal lines). For the terrestrial neutrino
beam, we assume vacuum oscillations (or short enough baselines) and a measurement at the
atmospheric oscillation maximum.
 }
\end{figure}

We show in \figu{sources} the exact dependence
of the observables $R$ (astrophysical neutrinos) and $P_{\mu e}$ (neutrino beams) on $\deltacp$ for different sources. 
While astrophysical sources have the largest modulation of the amplitude at $\deltacp=0$ and $\pi$ because of the $\cos \deltacp$-dependence, neutrino
beams are strongly influenced at $\deltacp = \pm \pi/2$ because of the  $\sin \deltacp$-dependence
at the oscillation maximum. In addition, as discussed above, neutrino beams show a different behavior for the neutrino and antineutrino operation modes. However, note that the comparison between neutrinos and antineutrinos can do only very little close to $\deltacp=0$ and $\pi$. 

In order to illustrate the measurement precision of $\deltacp$, we show possible error bars for $5\%$, $10\%$, and $20 \%$ measurement errors as the shaded bars. From the projection of the curves onto these bars, we can immediately read off the required precisions for $R$ and the relevant parameter regions in $\deltacp$ (for large $\stheta$) for the different sources: For neutron beam sources, we may learn something 
about $\deltacp$ already for a 20\% error on $R$ -- provided that $\deltacp$ is close to $0$ or $\pi$.
If considerably higher precisions can be achieved, other values of $\deltacp$ can be extracted as well. For muon damped sources, the required precision in $R$ is similar. For pion sources, however, only precisions of the order of $5\%$ can help.

\begin{figure}[t!]
\begin{center}
\includegraphics[width=8cm]{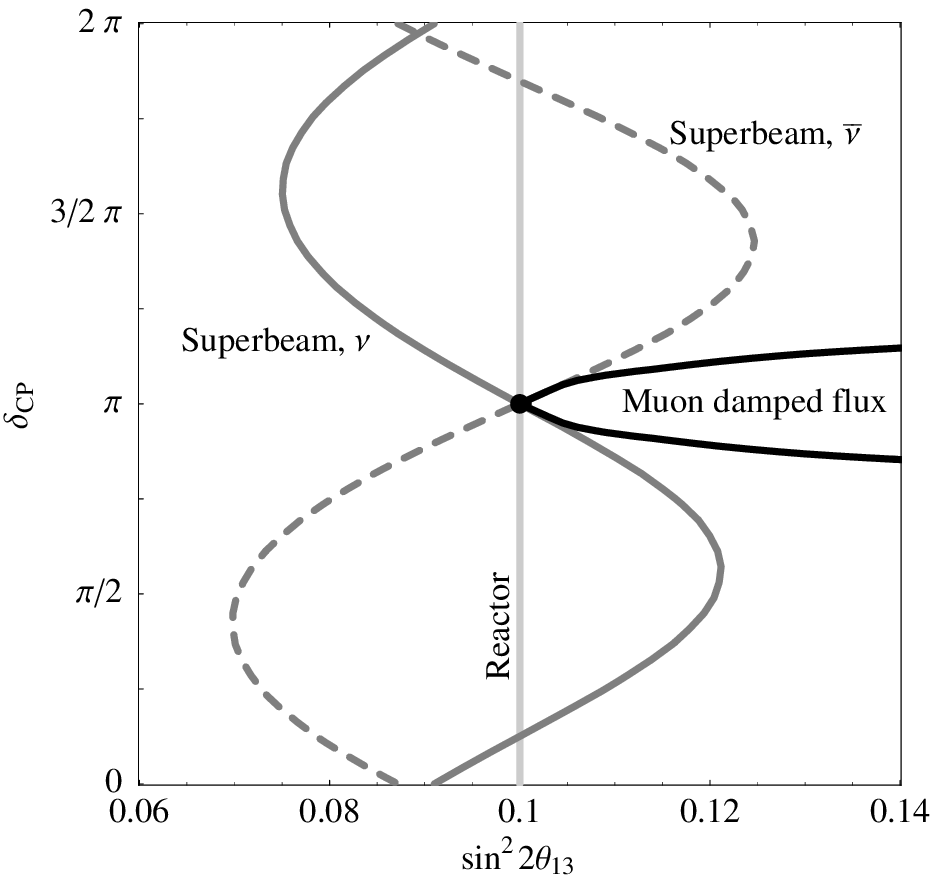}
\end{center}
\mycaption{\label{fig:didrates} Illustration of the synergy among superbeams, reactor experiments, and astrophysical
fluxes (at the example of a muon damped source) in $\stheta$-$\deltacp$-space. Shown are the curves for constant rates
(superbeam, reactor experiment) and constant R (astrophysical flux) going through the
best-fit point $\stheta=0.1$ and $\deltacp=\pi$. 
 }
\end{figure}

A different way to look at the synergy among  superbeams, reactor experiments, and astrophysical
fluxes is illustrated in \figu{didrates}. In this figure, the curves for constant total rates
(superbeam, reactor experiment) and constant R (astrophysical flux) going through the
best-fit point $\stheta=0.1$ and $\deltacp=\pi$ are shown. For the superbeams, this means that without spectral information 
all the points on the respective curves are degenerate, which is a good approximation for narrow-band beams. 
The combination of neutrinos and antineutrinos still leaves a remaining degeneracy between $0$ and $\pi$, which can,
for large error bars, not be resolved by the reactor experiment either, because the  degenerate solutions are still
too close to each other. However, the astrophysical flux has a completely different dependence on $\stheta$ and $\deltacp$,
which means that it could disentangle the two values for $\deltacp$ ($0$ and $\pi$). Note that all the values on the reactor and astrophysical flux curves are exactly degenerate even when using energy information. 

As far as different measurements are concerned, we expect an impact of the astrophysical
sources on CP precision measurements (especially for $\deltacp$ close to $0$ and $\pi$), and
for the mass hierarchy measurements at the superbeams because the $\mathrm{sgn}(\ldm)$-degeneracy
is located at a different value of (fake) $\deltacp$ than the original solution. We have also 
tested the impact on $\stheta$ exclusion and $\stheta$ discovery potentials, we we have not
found any significant impact. The reason is that these measurements are dominated by the ratio
of signal to background in the appearance channels of the beams, \ie, the absolute event rate determines the performance. This event rate is, for neutrinos and a normal mass hierarchy, smallest close to $\deltacp \sim \pi/2$ (\cf, \figu{sources}, lower right panel), \ie, $\deltacp \sim \pi/2$ (or $3 \pi/2$ for the inverted hierarchy) limits the performance. Since $R$ for the astrophysical sources does not depend strongly on $\stheta$ close to $\deltacp = \pi$ or $3 \pi/2$, they are of little help (\cf, \figu{sources}, first three panels). Note that information on $\stheta$ cannot be expected from an astrophysical source alone either, because a different combination of $\deltacp$ and $\theta_{23}$ can easily fake $\stheta=0$.

Finally, note that $R$ depends on $\theta_{23}$ in a different way than $P_{\mu e}$ for the beams
(\cf, \Refs~\cite{Akhmedov:2004ny,Serpico:2005bs}): While $R$ is in some way proportional to
$\theta_{23}$ by different combinations of powers of $\sin \theta_{23}$ and $\cos \theta_{23}$, the main sensitivity for the neutrino beams comes from the
term $\propto \sin^2 \theta_{23} \, \theta_{13}^2$ term in $P_{\mu e}$. We will discuss this different
dependence in \Sec~\ref{sec:theta23}.

\section{Simulation method}
\label{sec:sim}

In order to simulate the impact of $R$ on the future terrestrial neutrino oscillation
program, we use a modified version of the GLoBES software~\cite{Huber:2004ka}. Let us define
the input parameter vector used for GLoBES as 
\begin{equation}
\overrightarrow{x} = (\theta_{12}, \theta_{13}, \theta_{23}, \deltacp, \sdm, \ldm, \rho_1, \hdots, \rho_n) \, ,
\nonumber
\end{equation}
 where we simulate the combination of $n$ terrestrial experiments with the
average matter densities $\rho_i$, \ie, we deal with $6+n$ parameters. Note that the matter densities
are treated as oscillation parameters in GLoBES in order to include correlations with 
matter density uncertainties. We implement the $\chi^2$ as function of the simulated
and fit parameter vectors as
\begin{equation}
\chi^2 = \chi^2_{\mathrm{Terr}} (\overrightarrow{x}_{\mathrm{sim}}, \overrightarrow{x}_{\mathrm{fit}}) + \frac{\left( R(\overrightarrow{x}_{\mathrm{sim}}) - R(\overrightarrow{x}_{\mathrm{fit}}) \right)^2}{\sigma_R^2} \, , 
\label{equ:chi2}
\end{equation}
where $\sigma_R = \sigma^{\mathrm{rel}}_R \cdot R(\overrightarrow{x}_{\mathrm{sim}})$ is the absolute error on $R$
determined from an astrophysical source, and $\sigma^{\mathrm{rel}}_R$ is the relative error which we
will be using. In this equation, $\chi^2_{\mathrm{Terr}}$ is the $\chi^2$ in GLoBES describing
the terrestrial experiments and external input {\em before} marginalization over the oscillation parameters, and $R$ is the flux ratio using the exact analytical expressions from \equ{p}. We marginalize $\chi^2$ in \equ{chi2} over any unused oscillation parameters as defined
by the specific performance indicator (such as over all parameters except from $\deltacp$ for a 
$\deltacp$ precision measurement). Note that $R$ does not depend on the mass squared differences
and matter densities, which means that only four of the $6+n$ parameters determine the second
term of \equ{chi2}.

As far as the time scale for this study is concerned, we require substantial information from
both a neutrino telescope and superbeam experiments. Therefore, we discuss neutrino oscillation
physics at the end of the first-generation superbeam era, which could be around ten years
from now (until about 2015-2017). At that time, we assume that data from MINOS, Double Chooz, T2K, and
NO$\nu$A will be available for the part of the terrestrial experiments. In addition, there might
be early large reactor experiments, such as the Double Chooz upgrade ``Triple Chooz''~\cite{Huber:2006vr},
which we do not consider since they are less established yet. We use the MINOS simulation
from \Ref~\cite{Huber:2004ug} with a total luminosity of
$5 \, \mathrm{yr} \times 3.7 \, 10^{20} \, \mathrm{pot}/\mathrm{yr}$ and a $5.4 \, \mathrm{kt}$ magnetized iron calorimeter~\cite{Ables:1995wq} (the unit ``pot/yr'' refers to ``protons on target per year'').
For Double Chooz, we use the simulation from \Ref~\cite{Huber:2003pm} applied to Double Chooz,
\ie, $L=1.05 \, \mathrm{km}$ and an integrated luminosity of $60 \, 000$ unoscillated events~\cite{Ardellier:2004ui}, corresponding to about three years of data taking with both
near and far detectors.
For T2K, we use the simulation from \Ref~\cite{Huber:2002mx} based upon
\Ref~\cite{Itow:2001ee} with a total luminosity of $5 \, \mathrm{yr} \times 1.0 \, 10^{21} \, \mathrm{pot}/\mathrm{yr}$ and a $22.5 \, \mathrm{kt}$ water Cherenkov detector with neutrino running only. For NO$\nu$A, we use the simulation from \Ref~\cite{Huber:2002rs} updated to the numbers from \Ref~\cite{Ayres:2004js}, \ie, $L=810 \, \mathrm{km}$ operated $12 \, \mathrm{km}$ off-axis with $3 \, \mathrm{yr} \times 6.5 \, 10^{20} \, \mathrm{pot}/\mathrm{yr}$ for neutrinos and $3 \, \mathrm{yr} \times 6.5 \, 10^{20} \, \mathrm{pot}/\mathrm{yr}$ for antineutrinos and a $30 \, \mathrm{kt}$ totally active scintillating detector (unless stated otherwise). In most cases, we
will show the combination of these experiments, unless we use the reactor experiment alone
for a somewhat shorter time scale. Note, however, that MINOS only has a marginal impact on
the final results (\ie, the finally achieved MINOS luminosity hardly matters), 
and the main results will be mainly determined by accumulated statistics from
the superbeams and Double Chooz, the mass hierarchy sensitivity of NO$\nu$A, and the ability of
Double Chooz to cleanly extract $\stheta$. From the conceptual point of view, we therefore
expect that the qualitative results will be hardly affected by the finally achieved 
luminosities as long as the latter experiments are present with sufficient statistics.

Except from the information coming from the simulated experiments, we 
assume an external precision of 5\% for each $\sdm$ and $\theta_{12}$. 
This should approximately correspond to the precision at the time of the
data analysis (see, \eg , \Ref~\cite{Bahcall:2004ut}). In addition,
we include a 5\% uncertainty on the value of the baseline-averaged 
matter density, where the uncertainty takes into account matter 
density uncertainties as well as matter density profile 
effects~\cite{Geller:2001ix,Ohlsson:2003ip}. Since a reactor experiment
needs some information on the leading atmospheric
parameters, we impose a 10\% external error on $\ldm$ for the analysis
of a reactor experiment alone. 

\section{Early knowledge on $\boldsymbol{\deltacp}$ from
reactor experiments?}
\label{sec:reactor}

\begin{figure}[t!]
\begin{center}
\includegraphics[width=0.8\textwidth]{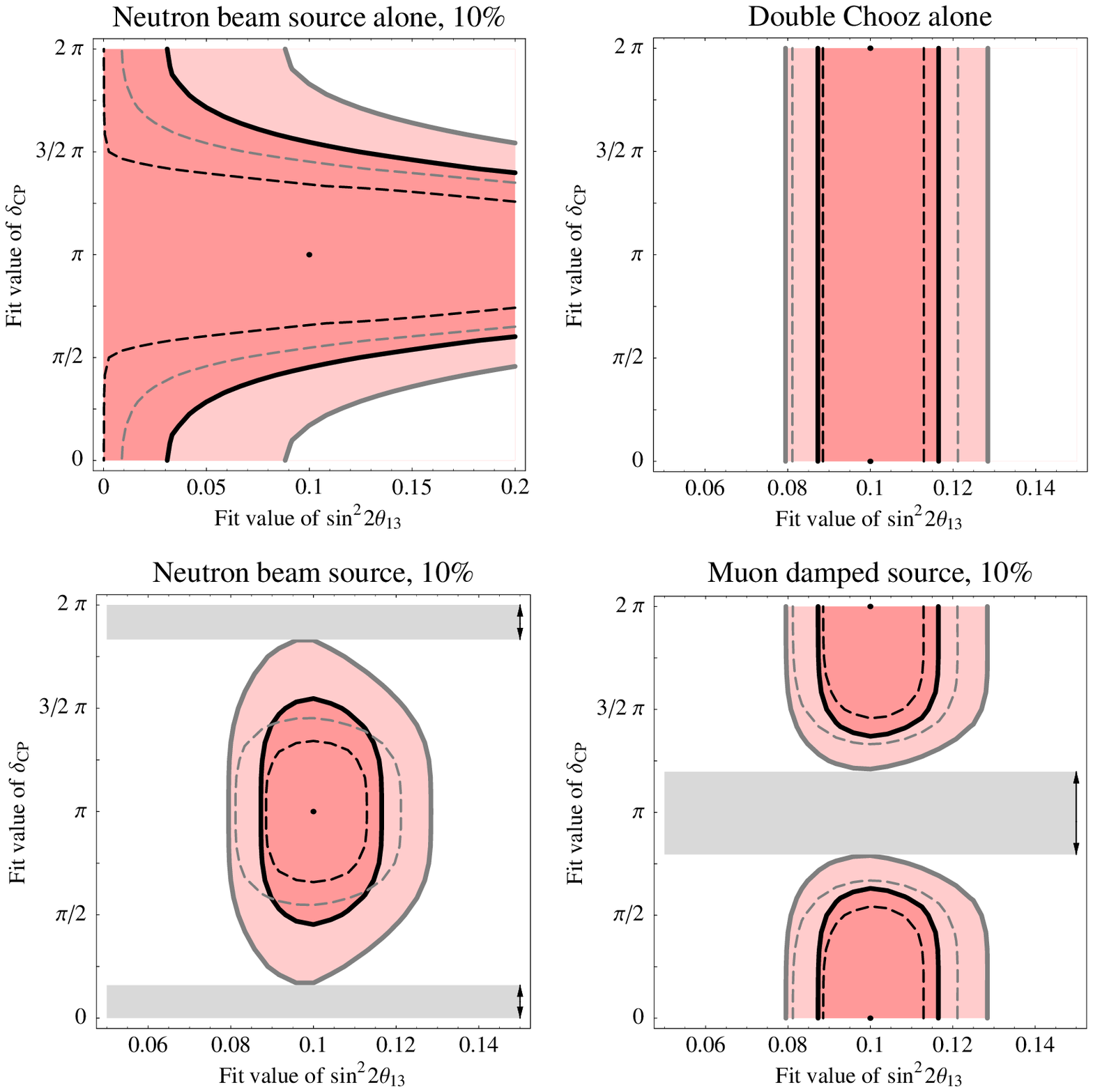}
\end{center}
\mycaption{\label{fig:reactor} Fit regions 
 as function of $\stheta$ and $\deltacp$ for different experiments as given in the plot captions
 (the lower row is always in combination with Double Chooz). The simulated values
 are chosen as  marked by the dots. The contours are shown for the $1\sigma$ (black curves, dark regions) and 90\% (gray curves,
 light regions) confidence level (1 d.o.f.). Dashed curves represent the results when the other (not shown) oscillation
 parameters are fixed, \ie, not marginalized over. The arrows in the lower row mark the ranges in $\deltacp$ which can be
 excluded at the 90\% confidence level.}
\end{figure}

The next generation of experiments providing some information on $\stheta$ should be reactor experiments. In particular,
Double Chooz could measure $\stheta$ already by about 2011 or so~\cite{Ardellier:2004ui} before the beams may have started running. Let us assume
that $\stheta$ is large, \ie, close to the current upper bound. In this case, Double Chooz could measure $\stheta$ relatively
precisely (\cf, upper right panel of \figu{reactor}). However, the results will not depend on $\deltacp$, which means that
it is impossible to obtain information on $\deltacp$ from Double Chooz alone.  In addition, it is difficult to obtain information on $\deltacp$ from an astrophysical source alone as well
after the other oscillation parameters have been marginalized over 
(\cf, upper left panel of \figu{reactor}). If, however, an astrophysical source is able
to provide this information on a similar timescale as the reactor experiment, 
one will actually be able to learn something on $\deltacp$ already before
the superbeams provide results. We show in \figu{reactor}, lower row, two examples for the combination of an astrophysical and Double
Chooz signal. In both of these cases with a 10\% precision measurement, a range of values of $\deltacp$ (arrows) could be excluded
at the 90\% confidence level. Of course, a CP violation measurement will not be possible because of the low precision and the
low sensitivity of the astrophysical sources close to maximal CP violation.

As far as the dependence on the true $\deltacp$ is concerned, $\deltacp = 0$ and $\pi$ for neutron beam and muon damped sources yield
similar qualitative results, For $\deltacp$ close to maximal CP violation, only muon damped and pion sources can provide some
hint for excluded values of $\deltacp$ at the $1 \sigma$ confidence level for high precisions of $R$ (about 5\%). If the precision of the 
astrophysical measurement is only 20\%, there will be some hints for some points at the $1\sigma$ confidence level for
neutron beam and muon damped sources. A $3\sigma$ exclusion is only possible for a muon damped source if $\deltacp=0$. In this case,
one can actually exclude $\deltacp=\pi$ at the $3\sigma$ confidence level.

\section{Mass hierarchy determination}
\label{sec:mh}

\begin{figure}[t]
\begin{center}
\includegraphics[width=8cm]{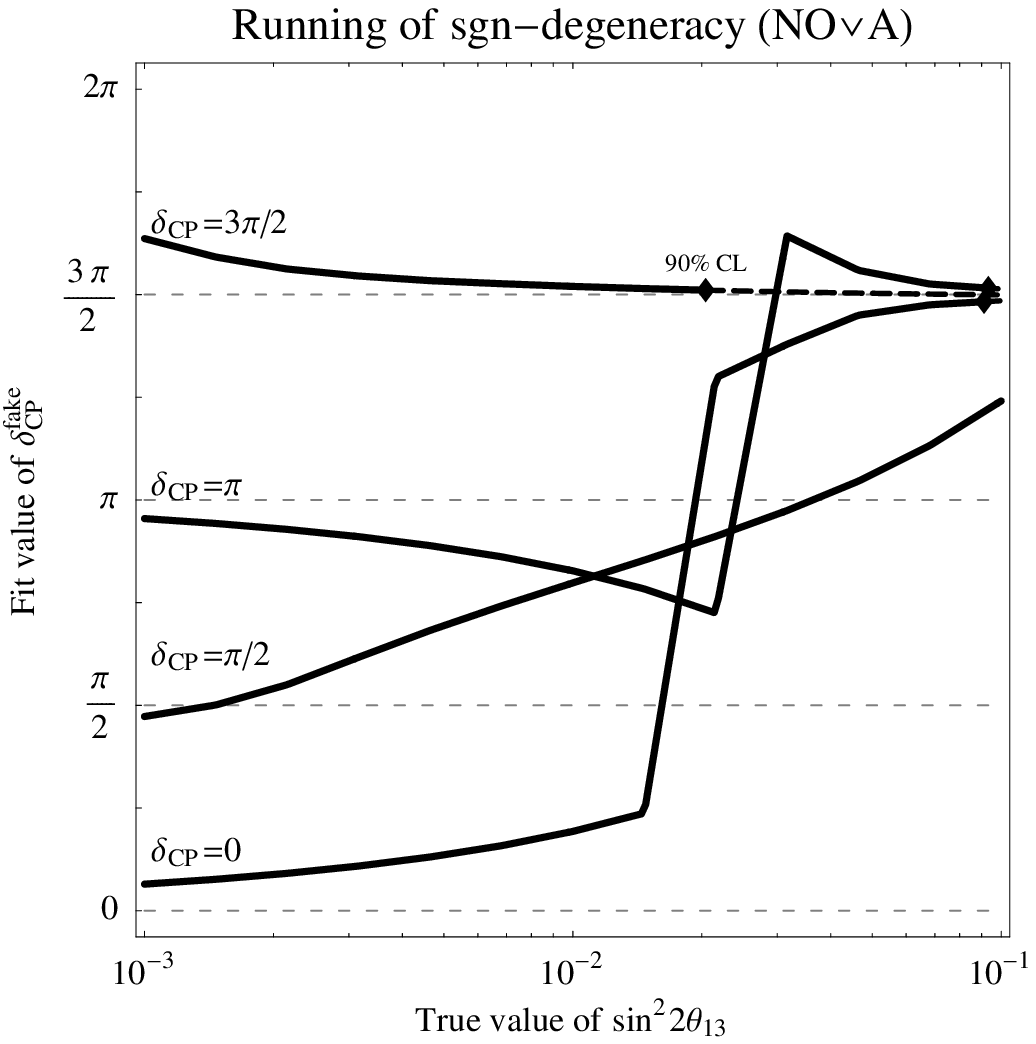}
\end{center}
\mycaption{\label{fig:degrun} Fake solution $\deltacp$ for the $\mathrm{sgn}(\ldm)$-degeneracy as function of the true $\stheta$ for different values of the true $\deltacp$ as given in the plot.
Diamonds mark the points beyond which (to the right) the degeneracy can be resolved at the
90\% confidence level. Figure for NO$\nu$A (3~yr neutrinos+3~yr antineutrinos) and a 
normal hierarchy.}
\end{figure}

The mass hierarchy determination using astrophysical flavor ratios alone
will not be possible because the observables do not depend on the mass hierarchy
for averaged oscillations. 
The first experiment from the next generation of beam
experiments with sufficiently large matter effects to determine
the mass hierarchy will be NO$\nu$A, \ie, the mass hierarchy sensitivity
will be dominated by NO$\nu$A. However, the $\mathrm{sgn}(\ldm)$-degenerate
solution of NO$\nu$A, in general, appears at a different value of $\deltacp$
than the original solution, which means that the correlations with $\deltacp$ and
$\stheta$ highly affects the mass hierarchy sensitivity. 
We illustrate in \figu{degrun} the running of the fake solution
in $\deltacp$-space for this experiment for different simulated values of $\deltacp$ as function
of the true $\stheta$. Note that the degeneracy can only be resolved on the right-hand sides
of the diamonds of the curves ($90\%$ CL). One can identify two sets of fixed points in this graph:
For small values of $\stheta$, the fake solution is close to the original one (or $\pi - \deltacp$). 
For very large $\stheta$, the fake solution is close to an attractor point which is, for this beam, $\deltacp=3 \pi/2$ for the
normal hierarchy and $\deltacp=\pi/2$ for the inverted hierarchy.\footnote{In order to
understand this dependence and attractor points, the concept of bi-probability/bi-rate graphs is
useful (\cf, \Refs~\cite{Minakata:2001qm,Minakata:2002qi,Minakata:2002qe,Winter:2003ye}). For small $\stheta$, the original and
fake solution ``pencils'' overlap completely, and the fake solution $\deltacp$ is close to the original $\deltacp$ or $\pi - \deltacp$ (\cf, \fig~2 in \Ref~\cite{Winter:2003ye}). For very large $\stheta$, where the pencils are separated, the fake solution is close to an attractor point determined by the closest point of the fake solution ellipse to the original solution ellipse (for this beam, $\deltacp=3 \pi/2$ for the normal hierarchy and $\deltacp=\pi/2$ for the inverted hierarchy). Note that for T2K matter effects are too small to separate the two pencils sufficiently, \ie, the addition of the astrophysical source would only shift the fake solution (still appearing at a low $\Delta \chi^2$).}
As it is obvious from \figu{degrun}, a constraint on $\deltacp$ would increase the confidence level of the degenerate
solution if it was strongly running, \ie, very different from the original solution. Therefore,
we expect an improved mass hierarchy sensitivity in combination with an astrophysical neutrino flux.

\begin{figure}[t]
\begin{center}
\includegraphics[width=\textwidth]{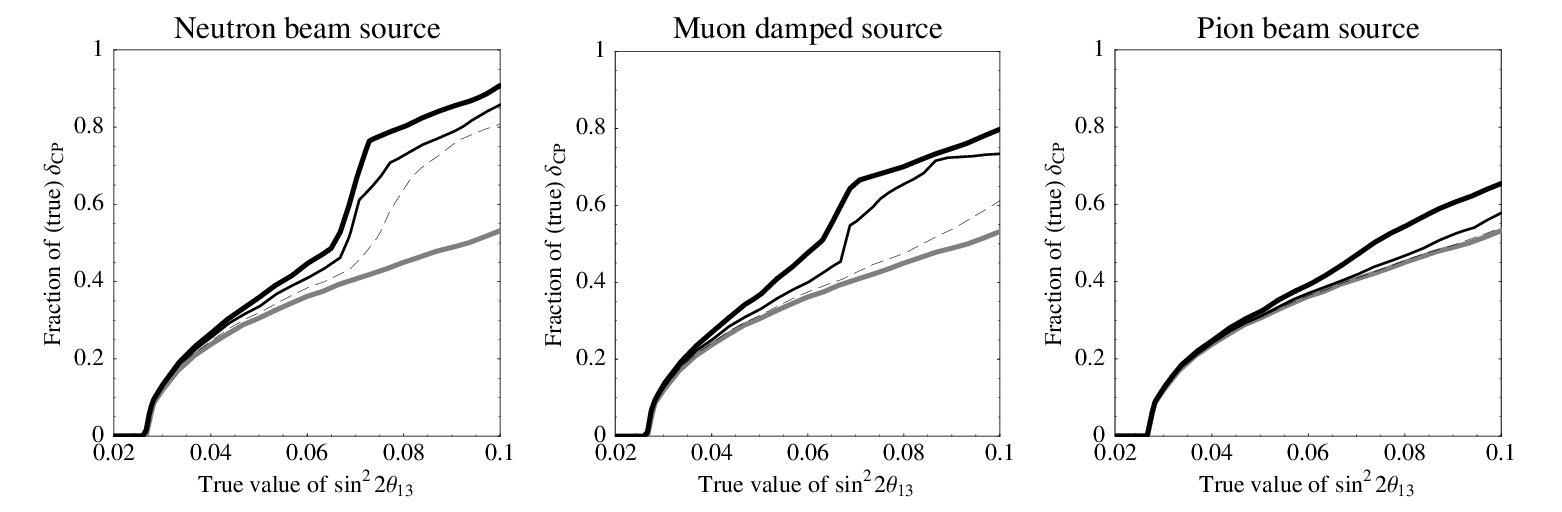}
\end{center}
\mycaption{\label{fig:massh} Sensitivity to the normal mass hierarchy as function of true $\stheta$ and true $\deltacp$ (stacked to the ``Fraction of $\deltacp$'') for MINOS, Double Chooz, T2K,
and NO$\nu$A combined with an astrophysical neutrino source (as given in the plot captions) at the $2 \sigma$ confidence level. The curves are for the following errors on the astrophysical flux ratio: no astrophysical flux observed (thick gray curves), 20\% error on $R$ (dashed curves), 10\% error on $R$ (thin black curves), and 5\% error on $R$ (thick black curves).}
\end{figure}

In order to test this hypothesis, we show in \figu{massh} the sensitivity to the normal mass hierarchy as function of true $\stheta$ and true $\deltacp$ (stacked to the ``Fraction of $\deltacp$'') for MINOS, Double Chooz, T2K,
and NO$\nu$A combined with an astrophysical neutrino source. The interpretation
of this figure is as follows: For large $\stheta$, the terrestrial experiments alone will only
be able to determine the mass hierarchy for about 50\% of all possible values of $\deltacp$
which could be realized by nature. However, using, for instance, a neutrino beam source
flux measured with a precision of 20\% increases this fraction to 80\% of all values of 
$\deltacp$. Therefore, depending on chosen confidence level and precision of the astrophysical
flux, the chance to discover the mass hierarchy will be improved from about half of all
possible cases of $\deltacp$ to almost certain. Similar results can be expected from a muon damped source, while for a pion beam source the fraction of $\deltacp$ will be increased by up to 10\%.
Note that while MINOS, T2K, and Double Chooz combined do not have any relevant sensitivity to
the mass hierarchy (\cf, \Ref~\cite{Huber:2004ug}), NO$\nu$A alone could not achieve this impressive result in combination with the astrophysical sources either. Only the combination of these experiments will provide the information to disentangle $\stheta$, $\deltacp$, and the mass hierarchy, as well
as it will have sufficient statistics.
In addition, we expect qualitatively similar results for the inverted hierarchy, where the role of $\deltacp = \pi/2$ and $3 \pi/2$ is exchanged for the superbeams, while the astrophysical sources still provide the relevant information close to $\deltacp=0$ and $\pi$.

\section{Measuring $\boldsymbol{\deltacp}$}
\label{sec:cp}

\begin{figure}[t]
\begin{center}
\includegraphics[width=\textwidth]{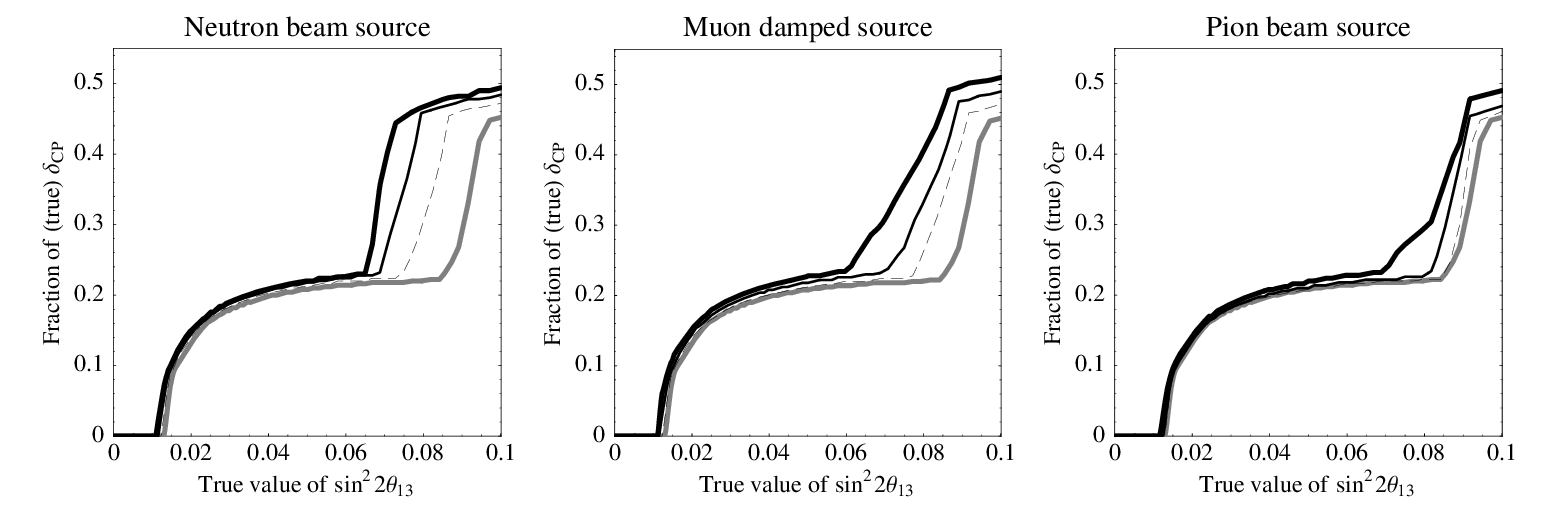}
\end{center}
\mycaption{\label{fig:cpviol} Sensitivity to CP violation as function of true $\stheta$ and true $\deltacp$ (stacked to the ``Fraction of $\deltacp$'') for MINOS, Double Chooz, T2K,
and NO$\nu$A combined with an astrophysical neutrino source (as given in the plot captions) at the $2 \sigma$ confidence level (normal mass hierarchy assumed). The curves are for the following errors on the astrophysical flux ratio: no astrophysical flux observed (thick gray curves), 20\% error on $R$ (dashed curves), 10\% error on $R$ (thin black curves), and 5\% error on $R$ (thick black curves).}
\end{figure}

Learning about $\deltacp$ has two aspects: First, we want to measure leptonic CP violation.
Second, we want to determine $\deltacp$ precisely. Note that these two options are not necessarily correlated: If, for example, $\deltacp$ is very close to (but not equal to) $0$ or $\pi$, we will certainly not be able to measure this small CP violation with the next generation of experiments.
However, we may still be able to learn something about $\deltacp$, such as we might exclude certain ranges. While superbeams are targeted towards CP violation, the sensitivity of the astrophysical sources should be best close to CP conservation. Therefore, we expect small effects for CP violation measurements, and larger effects for CP precision/exclusion measurements.

We show in \figu{cpviol} the sensitivity to CP violation as function of true $\stheta$ and fraction of $\deltacp$ for MINOS, Double Chooz, T2K,
and NO$\nu$A combined with an astrophysical neutrino source. Note that the overall fraction of possible values for which this measurement could be done is rather small, \ie, between about 20\% and 40\% for large $\stheta$. The jump from 20\% to 40\% comes from enough statistics to reduce the size of the $\mathrm{sgn}(\ldm)$-degenerate solution such that it does not overlap the CP conserving values anymore. Therefore, it is present depending on statistics and chosen confidence level (for instance, for NO$\nu$A alone, it is not present). An astrophysical flux measurement could not increase the fraction of $\deltacp$, for which CP violation is measurable, substantially (because it is most sensitive close to the CP conserving values), but it could increase the reach in $\stheta$ for about 20\% of all values of $\deltacp$. Note, however, that a similar effect could be achieved with slightly increased statistics.

\begin{figure}[t]
\begin{center}
\includegraphics[width=\textwidth]{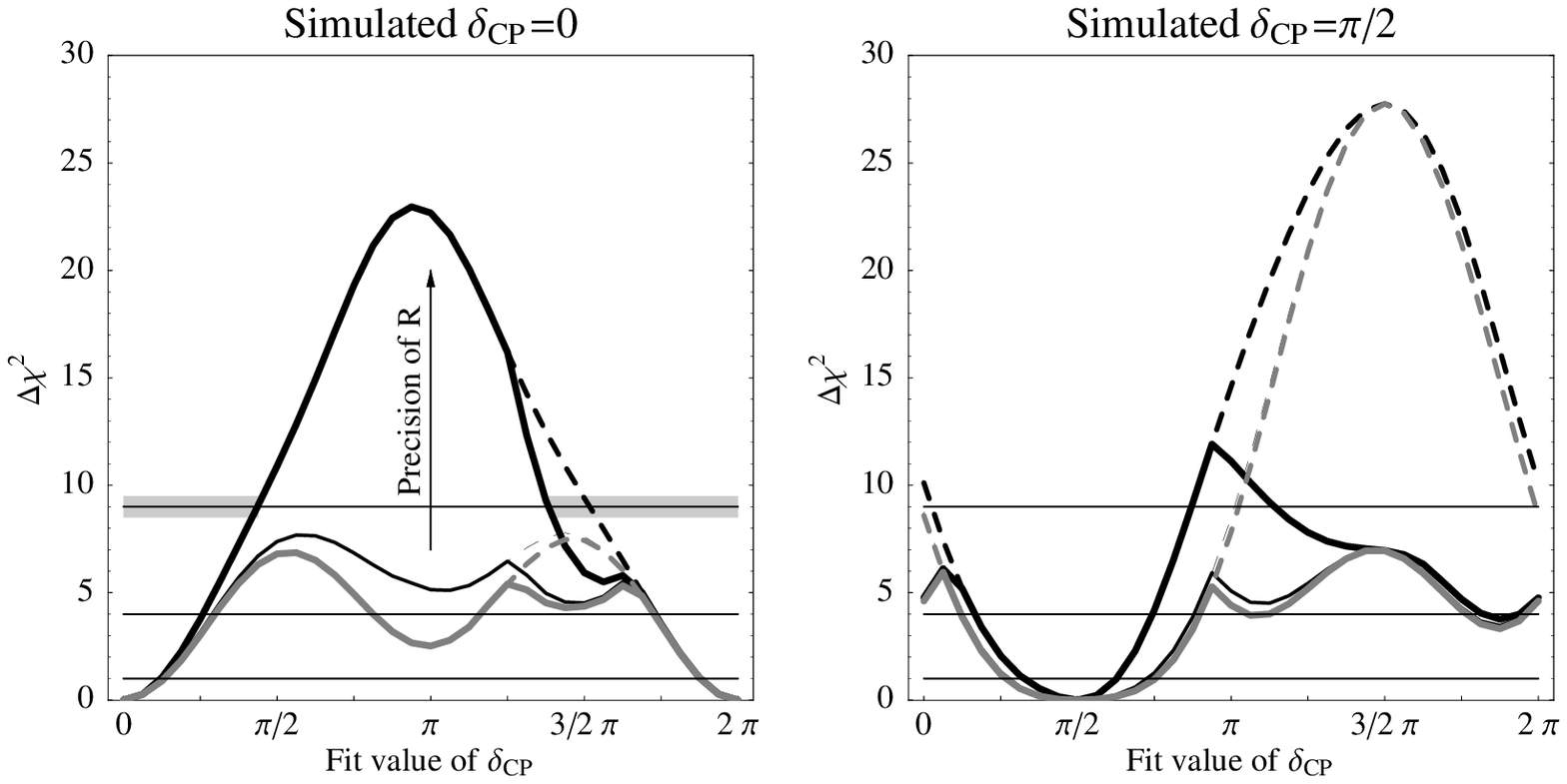}
\end{center}
\mycaption{\label{fig:didcpp} The $\Delta \chi^2$ as function of the fit value of $\deltacp$ for different simulated values of $\deltacp$ as given in the plot captions. The different curves correspond to no astrophysical information (thick gray curves), a 20\% measurement of $R$ from a muon damped flux (thin curves), and a 5\% measurement of $R$ from a muon damped flux (thick black curves), all in combination with the terrestrial experiments. The dashed curves would respresent the solutions if the mass hierarchy degeneracy was resolved. The CP coverage is obtained as summed fit ranges at a specific confidence level, such as illustrated for the black curve and $\Delta \chi^2=9$ in the left plot (gray bars).
}
\end{figure}

CP precision measurements are somewhat more complicated to describe, especially since the discussed experiments rather exclude some ranges of $\deltacp$ than they allow for a precision determination. For illustrative purposes, we show in \figu{didcpp} the $\Delta \chi^2$ as function of the fit value of $\deltacp$ for two different simulated values. The different curves correspond to no astrophysical information (thick gray curves), a 20\% measurement of $R$ from a muon damped flux (thin curves), and a 5\% measurement of $R$ from a muon damped flux (thick black curves), all in combination with the terrestrial experiments. For $\deltacp=0$ (left plot), the impact of the astrophysical source is obvious: the $\Delta \chi^2$ for the solution at $\deltacp=\pi$ is continuously lifted with increasing precision of $R$ (arrow). Note that for a 5\% precision measurement of $R$ a precision of $\deltacp$ is reached which allows to exclude maximal CP violation ($\deltacp=\pi/2$ and $3\pi/2$) almost at the $3 \sigma$ confidence level.
For $\deltacp=\pi/2$ (right plot), however, there is hardly any impact until high precisions of $R$ are reached, when the $\Delta \chi^2$ at $\deltacp=\pi$ is significantly increased. One can easily read off this effect from \figu{sources} (upper right): At $\deltacp = \pi/2$ the error bars have to be about a factor of two smaller than at $\deltacp = 0$ or $\pi$ to exclude some ranges in $\deltacp$.
As for the CP violation measurement above, the effect at the $2\sigma$ confidence level is small and the $\mathrm{sgn}(\ldm)$-degeneracy is already lifted above this confidence level at $\deltacp=\pi$ (the difference between the dashed and solid curves describes the impact of this degeneracy).

\begin{figure}[t]
\begin{center}
\includegraphics[width=\textwidth]{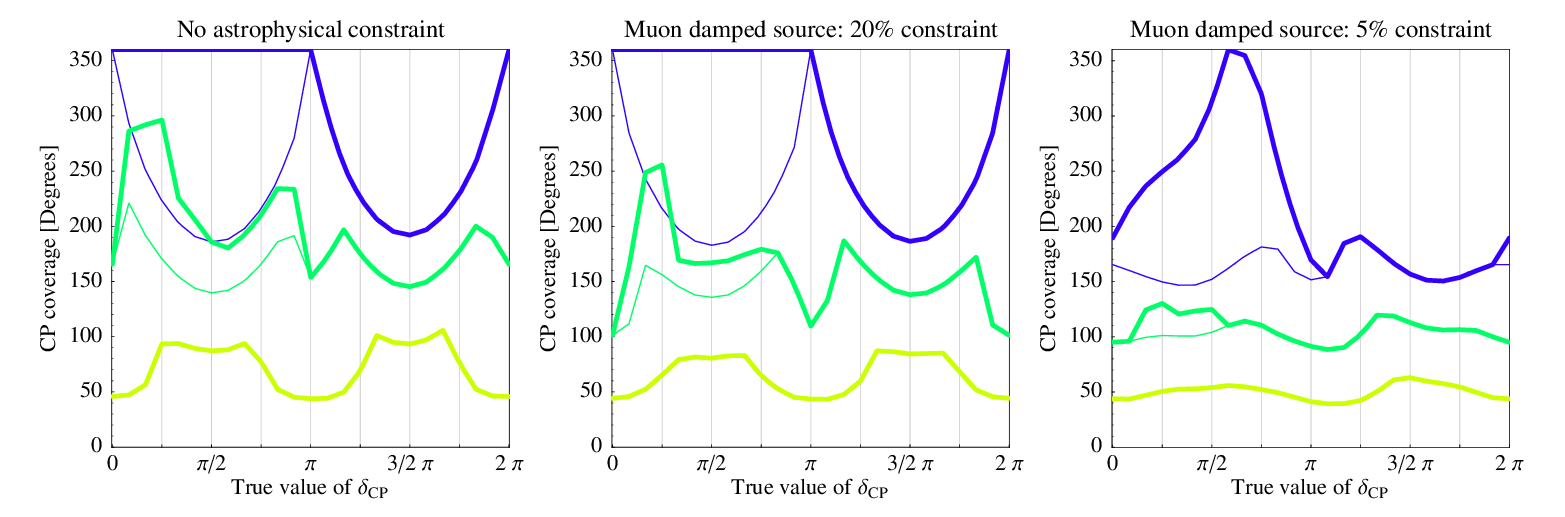}
\end{center}
\mycaption{\label{fig:cppat} CP patterns for MINOS, Double Chooz, T2K, and NO$\nu$A combined
for several external precisions on $R$ from an astrophysical muon damped source. The CP patterns quantify the measurement of $\deltacp$ (CP coverage) as a function of the true value of $\deltacp$ 
(provided by nature) for $\stheta=0.1$. The CP coverage is defined as range of fit values of $\deltacp$ which fit the chosen true value, and can be between $0$ (precise determination of
$\deltacp$) and $360^\circ$ (no information on $\deltacp$). The thick curves
correspond (from dark to light) to $\Delta \chi^2 = 9$, $4$, and $1$ respectively, and the thin curves represent the results without taking into account the $\mathrm{sgn}(\ldm)$-degeneracy.
}
\end{figure}

In order to study the CP exclusion/precision information as function of the simulated $\deltacp$, we introduce the CP coverage~\cite{Huber:2004gg}. The CP coverage is defined as range of fit values of $\deltacp$ which fit the chosen true value at a certain $\Delta \chi^2$, and lies between $0$ (precise determination of $\deltacp$) and $360^\circ$ (no information on $\deltacp$). It is illustrated in \figu{didcpp} as the gray bar in the left plot for the 5\% precision measurement of $R$ and $\Delta \chi^2=9$. We use in \figu{cppat} the ``CP patterns''~\cite{Winter:2003ye}, which quantify the measurement of $\deltacp$ (CP coverage) as a function of the true value of $\deltacp$ provided by nature.
This means that we compute \figu{didcpp} for all simulated values of $\deltacp$, read off the CP coverage from each figure, and show this performance indicator as function of the simulated $\deltacp$. \figu{cppat} shows these CP patterns for MINOS, Double Chooz, T2K, and NO$\nu$A alone, as well as combined with a muon damped source for $\stheta=0.1$ close
to the CHOOZ bound~\cite{Apollonio:1999ae}. Obviously, how precisely one can measure $\deltacp$ or what range of values one could exclude strongly depends on the chosen confidence level.\footnote{Note that strictly speaking, one should refer to the $\Delta \chi^2$ instead of the confidence level, since the connection between the shown $\Delta \chi^2$'s and the $1\sigma$, $2 \sigma$, and $3\sigma$ CL for 1 d.o.f. holds only for quantities with Gaussian errors.} However, it is obvious from this figure that astrophysical sources could especially help in the first two quadrants where the degeneracy
problem is present (for a normal hierarchy, for the inverted hierarchy it is the third and fourth quadrants), and close to $0$ and $\pi$. 
The CP patterns for a neutron beam source are qualitatively similar, whereas the pion beam source requires higher precisions.

The structure of the patterns in \figu{cppat} is highly non-trivial. In particular, for the terrestrial experiments only, the precision is best close to $\deltacp=0$ and $\pi$ at the $1\sigma$ confidence level, and worst at these points at $3 \sigma$. That is because the precision of $\deltacp$ depends very much on the size of the error bars. Going back to \figu{sources} (lower right) and studying the
impact of the error bars (shaded bars, terrestrial neutrino beams), one can read off that for small errors, the precision will be best close to $0$ and $\pi$ as $P_{\mu e}$ leaves the simulated value very quickly. For large errors of the order of the modulation amplitude, however, the precision will be worst close to $0$ and $\pi$ because at the values around the center easily all values of $\deltacp$ (for neutrinos and antineutrinos) can be covered (for more details, see \Ref~\cite{Winter:2003ye}). Note that the size of the error bars depends on statistics as well as the chosen confidence level. Therefore, one expects a behavior strongly dependent on the confidence level for CP precision measurements (see, \eg, \Ref~\cite{Huber:2004gg}). This argument can be translated to the astrophysical sources, where it will occur for $0$ and $\pi$ exchanged with $\pm \pi/2$. Therefore, we can understand this very complementary 
information from these two data sources. 

\begin{figure}[t]
\begin{center}
\includegraphics[width=\textwidth]{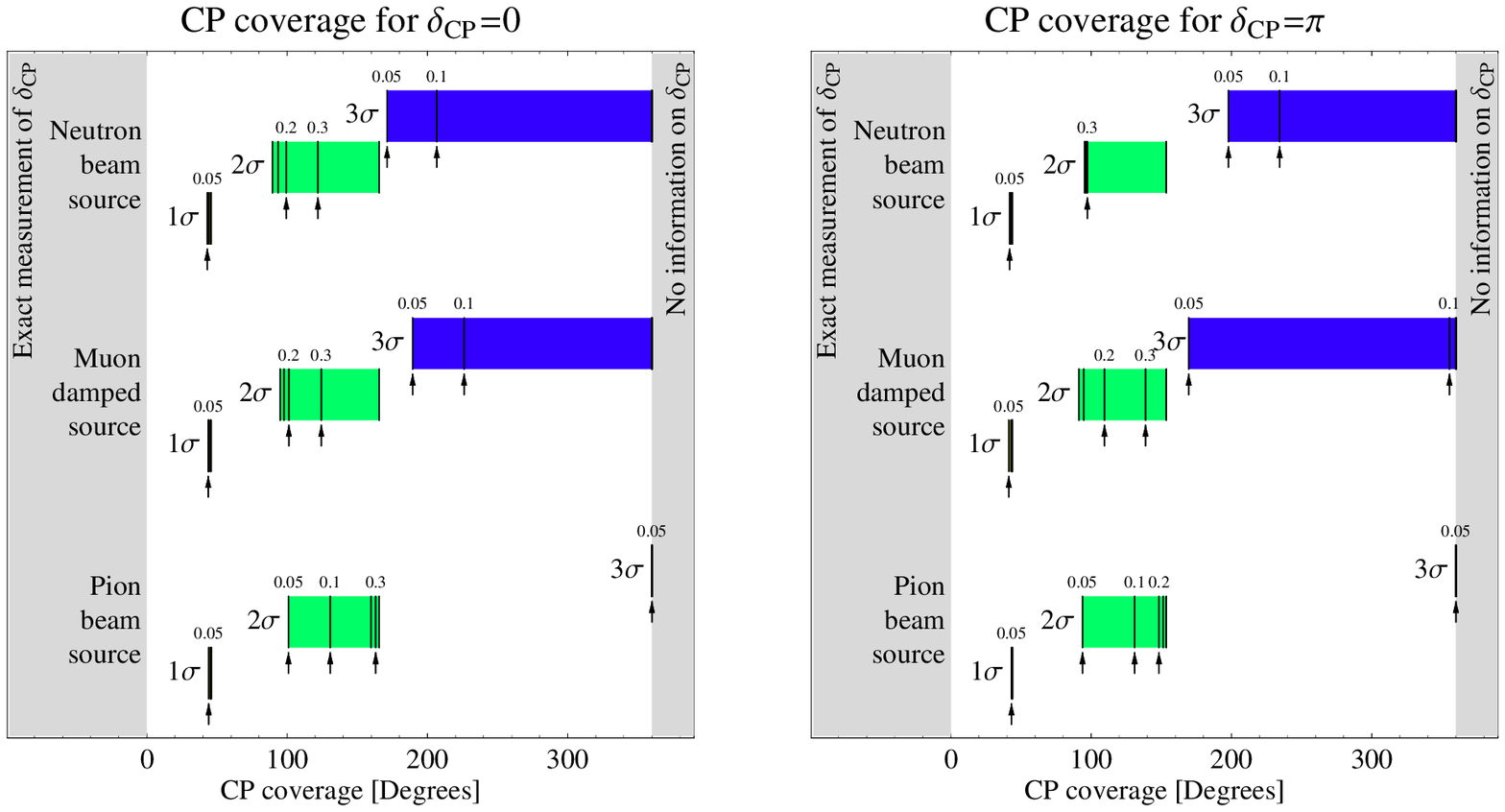}
\end{center}
\mycaption{\label{fig:cpbars} Impact of an astrophysical flux ratio measurement on the CP coverage for 
MINOS, Double Chooz, T2K, and NO$\nu$A combined for the true $\deltacp=0$ (left) and $\pi$ (right),
as well as $\stheta=0.1$. The bars represent the $1\sigma$, $2 \sigma$, and $3\sigma$ measurements ($\Delta \chi^2 = 1$, $4$, $9$) for different astrophysical sources as given in the plots. The right edges of the bars correspond to no astrophysical information, whereas the left edges correspond to a $5\%$ measurement if the respective flux ratio. The vertical lines in the bars represent (from right to left) no astrophysical information, a 30\% precision, a 20\% precision, a 10\% precision, and a 5\% precision, where only some of the lines
are labeled.
}
\end{figure}

In order to compare the confidence level dependence and dependence on the astrophysical source
more quantitatively, we show in \figu{cpbars} the impact of different astrophysical flux ratio measurements on the CP coverage for the true values of $\deltacp=0$ (left) and $\pi$ (right).
In this figure, the different vertical lines in the bars correspond to $5\%$, $10\%$, $20\%$, and $30\%$ measurements of $R$ (from left to right), as well as to no astrophysical measurement (right edges of bars). Selected precisions are marked with arrows and numbers. The results depend
very much on the confidence level, value of $\deltacp$, and flux used. However, one can
read off this figure that even a 30\% measurement of a neutron beam or muon damped flux can
have a substantial impact ($2 \sigma$, $\deltacp =0$). In addition, a 10\% precision of these
fluxes would allow for an exclusion of certain ranges at $3 \sigma$. Therefore, an astrophysical flux observation could be very useful to help restrict $\deltacp$.

\section{Resolving the octant degeneracy}
\label{sec:theta23}

\begin{figure}[t]
\begin{center}
\includegraphics[width=\textwidth]{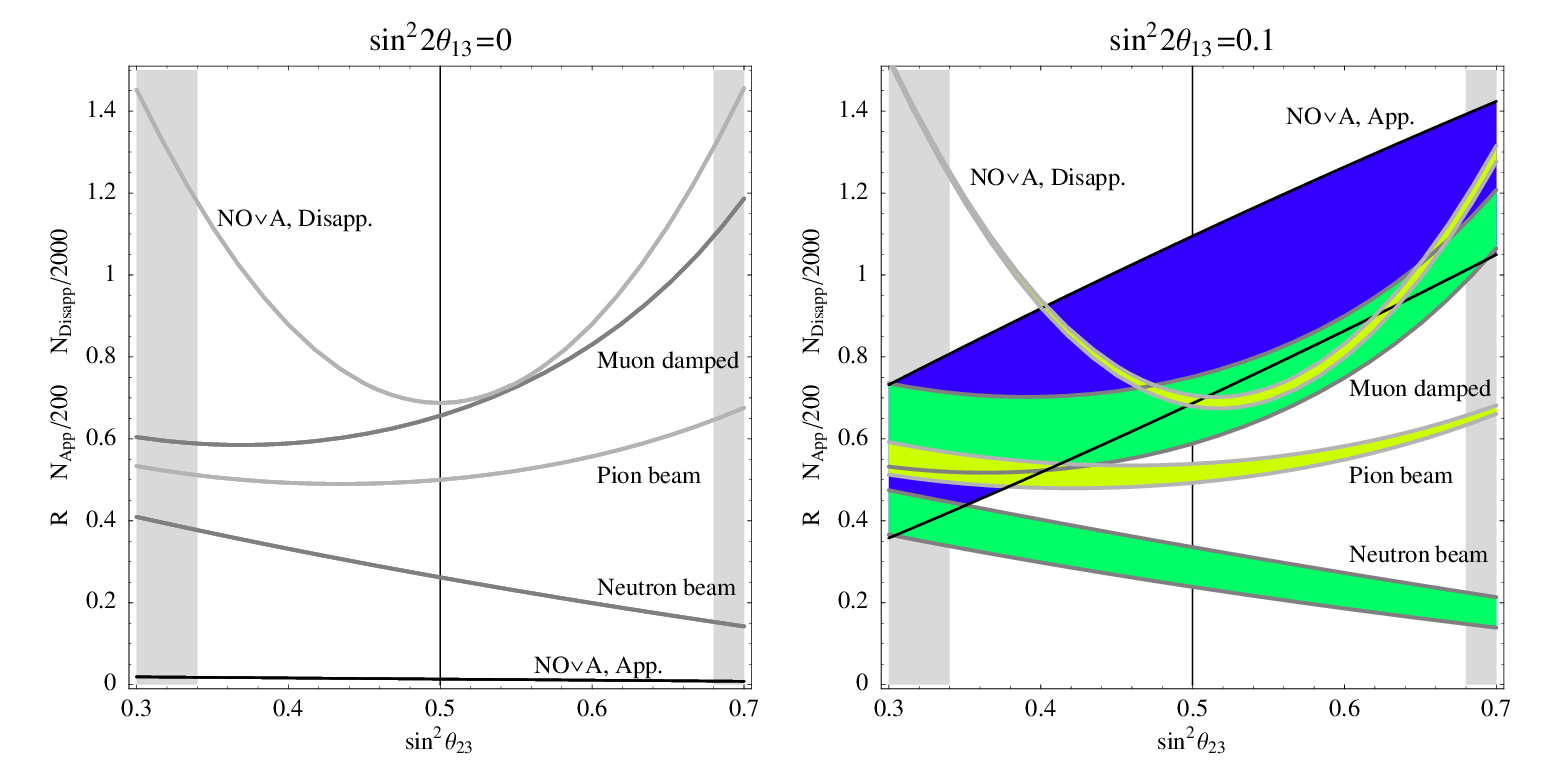}
\end{center}
\mycaption{\label{fig:didtheta23} Illustration of the observables ($R$ for astrophysical
sources, total event rates for the beam) for the octant degeneracy resolution 
as function of $\sin^2 \theta_{23}$ for $\stheta=0$
(left) and $\stheta=0.1$ (right). The bands reflect the unknown value of $\deltacp$. Note the
scaling of the vertical axes depending on the source considered. The gray-shaded areas 
mark the $3 \sigma$ excluded region~\cite{Maltoni:2004ei}. For NO$\nu$A, we assume five years
of neutrino running for this figure.
}
\end{figure}

As discussed in \Ref~\cite{Serpico:2005bs}, the flux ratio $R$ at neutrino telescopes has a distinctive dependence on $\theta_{23}$ which is sensitive to the $(\theta_{23},\pi/2 - \theta_{23})$-degeneracy.
Note, however, that without additional knowledge
on the other mixing parameters, this information can only be extracted from $R$
in very specific cases, but in a rather model-independent way~\cite{Serpico:2005bs}. We demonstrate in this section that the complementarity among superbeams, reactor experiments, and astrophysical sources allows for an exclusion of the octant degeneracy for any substantial deviation from maximal mixing.
For illustration, we show in \figu{didtheta23} the observables ($R$ for astrophysical
sources, total event rates for the beam) for the octant degeneracy resolution 
as function of $\sin^2 \theta_{23}$ for $\stheta=0$
(left) and $\stheta=0.1$ (right). The bands reflect the unknown value of $\deltacp$.
In order to discuss the potential to resolve the octant degeneracy, compare the left branch
of each source ($\sin^2 \theta_{23}<0.5$) with the right branch ($\sin^2 \theta_{23}>0.5$).
If we assume that a reactor experiment determines $\stheta$ fairly well and the impact of the unknown $\deltacp$ is one of the main uncertainties, this picture should be quite accurate.
For $\stheta=0$, the situation is quite simple because, in this limit, $\deltacp$ is meaningless.
The superbeam disappearance channel measures $\sin^2 2 \theta_{23}$ very precisely, but it does not have information on the octant. The rates in the appearance channel are too low to imply any information, and reactor experiment is not affected by $\theta_{23}$ (and therefore not shown).
However, the flux ration $R$ of the astrophysical sources is very different for $\sin^2 \theta_{23}<0$ and $\sin^2 \theta_{23}>0$, which means that a resolution of the octant degeneracy should be possible. However, for large $\stheta$, the situation is different: The uncertainties on
$\deltacp$ and $\stheta$ (not shown) should affect the astrophysical sources. The beam, on the other hand, now has a substantial appearance rate with some information on the octant. We therefore conclude that the measurement for small $\stheta$ should be dominated by the astrophysical source, whereas the measurement for large $\stheta$ could be dominated by the beam (plus reactor experiment). The latter hypothesis needs to be quantified, because it it unclear from \figu{didtheta23} how the correlations with the other oscillation parameters affect the degeneracy.

\begin{figure}[t]
\begin{center}
\includegraphics[width=8cm]{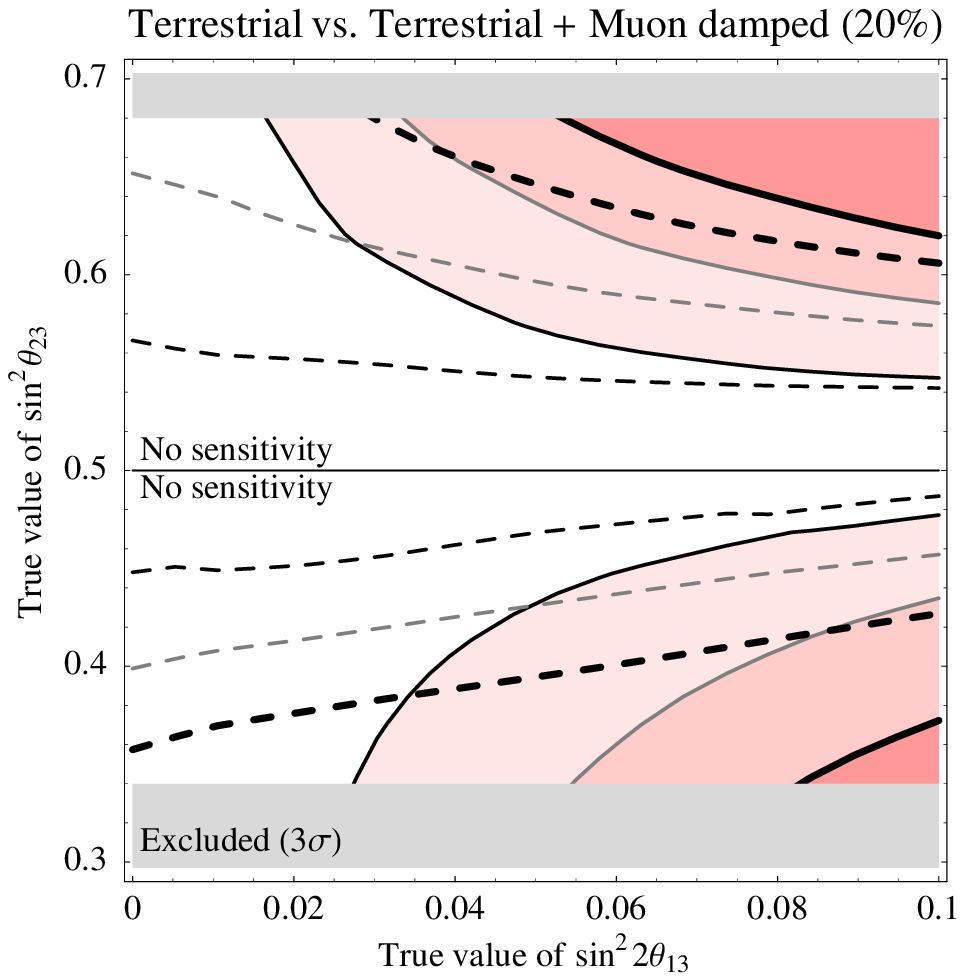}
\end{center}
\mycaption{\label{fig:theta23deg} Exclusion of the octant degeneracy as function of $\stheta$ and $\sin^2 \theta_{23}$ at the $1\sigma$ (thin black curves),
$2 \sigma$ (thin gray curves), and $3 \sigma$ (thick black curves) confidence level. The shaded areas
shows the results from the terrestrial experiments only (MINOS, Double Chooz, T2K, and NO$\nu$A combined), whereas the dashed curves represent the combination with a 20\% precision measurement
of a muon damped flux. The gray-shaded areas 
mark the $3 \sigma$ excluded region~\cite{Maltoni:2004ei}. Figure for $\deltacp=0$ and a normal
mass hierarchy.
}
\end{figure}

In order to demonstrate the resolution of the octant degeneracy quantitatively, we show in
\figu{theta23deg} the exclusion of the octant degeneracy as function of $\stheta$ and $\sin^2 \theta_{23}$. The shaded areas
show the results from the terrestrial experiments only (MINOS, Double Chooz, T2K, and NO$\nu$A combined), whereas the dashed curves show the combination with a 20\% precision measurement
of a muon damped flux as a representative example. Note that, for instance, NO$\nu$A alone could not resolve the octant degeneracy because of the correlation with $\stheta$ and $\deltacp$. However, it is the combination between NO$\nu$A and Double Chooz which allows for the degeneracy resolution for large $\stheta$ (shaded contours). As expected, the terrestrial experiments alone cannot resolve the degeneracy
for small $\stheta$, while the combination with a muon damped source (or similarly a neutron beam source) can. Note that this effect for small $\stheta$ is qualitatively comparable to the combination of atmospheric and long-baseline data using a megaton-size Water-Cherenkov detector~\cite{Huber:2005ep,Campagne:2006yx}, while it may be available at a shorter time scale. As far as the dependence on the simulated $\deltacp$ is concerned, we observe similar small qualitative effects
as in \Ref~\cite{Huber:2005ep}. 

\begin{figure}[t]
\begin{center}
\includegraphics[width=\textwidth]{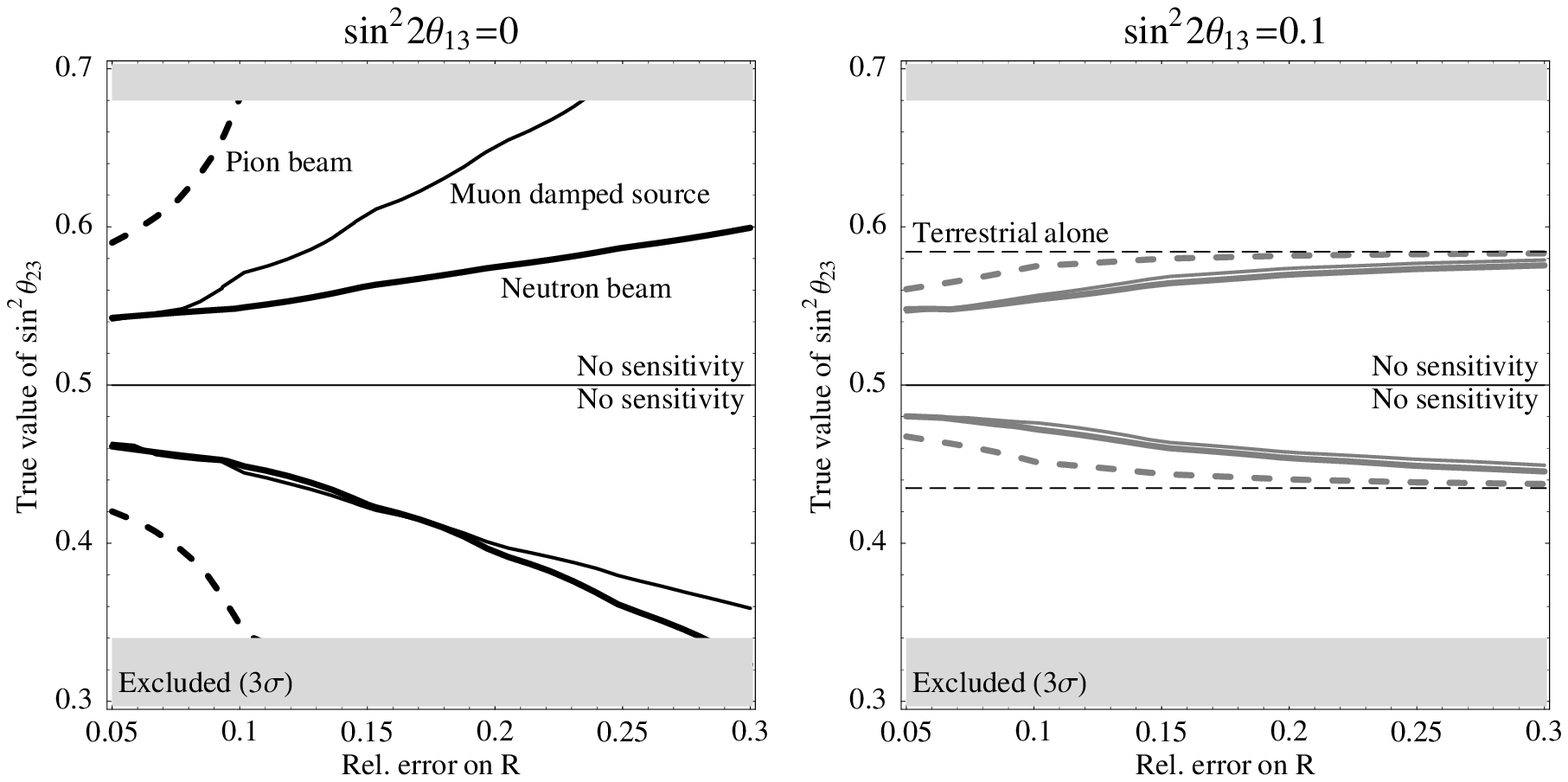}
\end{center}
\mycaption{\label{fig:theta23rdep} Exclusion of the octant degeneracy (above upper and below lower curves) as function of the precision of $R$ and $\sin^2 \theta_{23}$ at the $2\sigma$ confidence level.
The left plot corresponds to small (simulated) $\stheta=0$, whereas the right plot is shown for (simulated) $\stheta=0.1$.
The different curves are computed for the combination of different fluxes (neutron beam: thick curves; muon damped source: thin curves; pion beam: dashed curves) with the terrestrial experiments.
In the right plot, the result from the terrestrial experiments alone is shown as well. The gray-shaded areas  mark the $3 \sigma$ excluded region~\cite{Maltoni:2004ei}. Figure for $\deltacp=0$ and a normal mass hierarchy.}
\end{figure}

Since the dependence on $\stheta$ is quite straightforward for all astrophysical sources,
we focus in \figu{theta23rdep} on the two physics cases $\stheta=0$ (small) and $\stheta=0.1$ (large).
The figure shows the dependence of the octant resolution on the relative error on $R$ for different astrophysical sources. In the case $\stheta=0$, where the terrestrial experiments alone do not provide any substantial information, a 10\% to 20\% precision of $R$ from a neutron beam or muon damped flux would be very useful for a 10\% deviation from maximal mixing. For a pion beam source, however, the precision has to be very high. Note that even for poor measurements of a neutron beam flux some information for significant deviations from maximal mixing (but still within the currently allowed $3 \sigma$ range) could be obtained. For large $\stheta$, a 10\% to 20\% precision of $R$ from a neutron beam or muon damped flux could improve the reach in $\theta_{23}$ by a factor of two compared to the terrestrial experiments alone. However, beyond that the effect saturates quickly and the terrestrial experiments dominate the potential to resolve the degeneracy completely. Therefore, an astrophysical 
flux will be especially useful for small values of $\stheta$.
Note that, compared to \Ref~\cite{Huber:2005ep}, we have not considered mixed degeneracies (wrong octant and wrong mass hierarchy) in this study, but we expect that one could exclude this
degeneracy as well if the octant and wrong hierarchy alone can be excluded (similar to \Ref~\cite{Huber:2005ep}). In addition, we have only considered the exclusion of the octant degeneracy, which is much more difficult than the exclusion of maximal mixing (which can be done with the disappearance channel at beams). However, since the exclusion of the octant degeneracy is sufficient
to exclude maximal mixing (but not necessary), we do know that there will be some sensitivity to deviations from maximal mixing as well. We expect this to be much weaker than the one coming from the disappearance channels of the beams via the $\sin^2 2 \theta_{23}$-dependence independent of $\stheta$.

\section{What if one can actually measure all flavors?}
\label{sec:flavors}

So far, we have assumed that we cannot discriminate $\nu_e$ and $\nu_\tau$ shower events and we have
used $R= \phi_{\mu}/(\phi_{e}+\phi_{\tau})$ as an observable.
However, in principle, electromagnetic (from $\nu_e$) and hadronic showers (from $\nu_\tau$) can
be distinguished by their muon content. In addition, for large enough energies, the tau track
can be separated from the shower and may be identified by following 
double-bang and lollipop events (see \Refs~\cite{Beacom:2003nh,Jones:2003zy} and references therein). Since we do not know the flux normalization, we can describe the neutrino oscillation physics by two independent quantities completely. We choose $R$, as before,
and $S \equiv \phi_e/\phi_{\tau}$ which could be extracted from the ratio of electromagnetic to hadronic shower events.\footnote{Note that neutral currents produce hadronic showers, too. In this case, we refer to the hadronic showers from $\nu_\tau$ events only.} We expect that this quantity might be determined with a lower precision than $R$, and it only becomes an observable for higher energies. This could make it an interesting additional observable for astrophysical pion beam sources.

\begin{figure}[t!]
\begin{center}
\includegraphics[width=\textwidth]{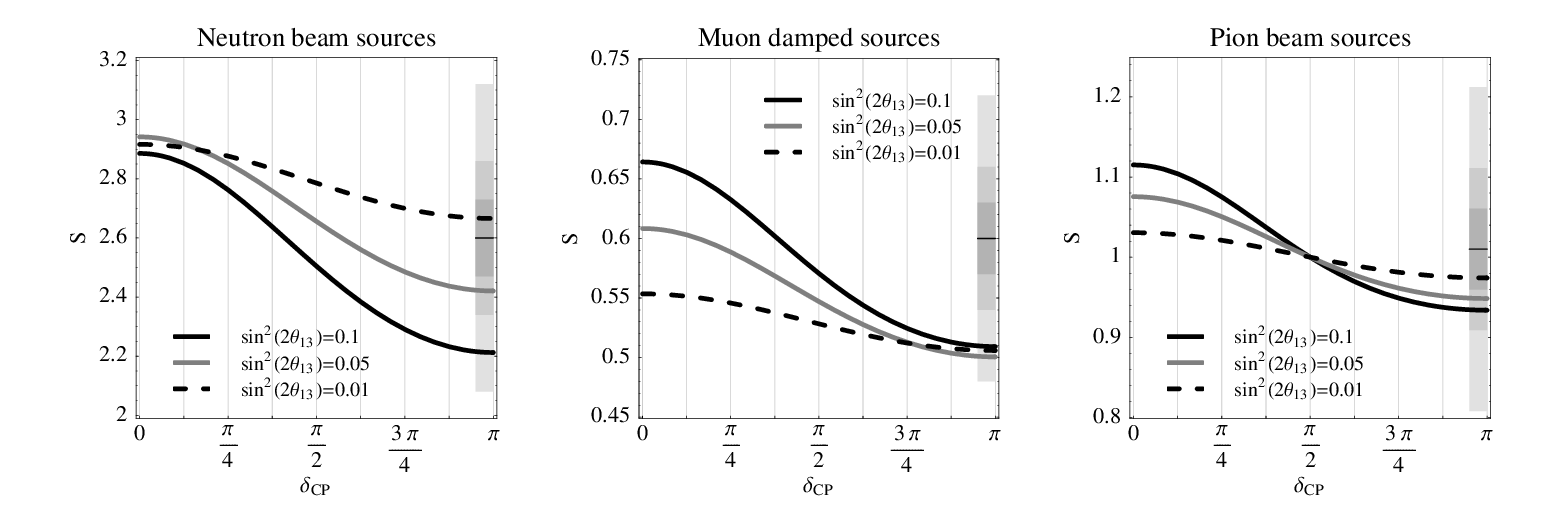}
\end{center}
\mycaption{\label{fig:ssources} The ratio $S=\phi_e/\phi_\tau$ (representing the electromagnetic versus hadronic shower events) 
as function of $\deltacp$ for the astrophysical sources. The shaded bars illustrate the size of the $5\%$, $10\%$, and $20\%$ errors for the chosen central values (horizontal lines). 
 }
\end{figure}

In order to discuss the qualitative dependence of the observable $S$ on $\deltacp$, we show it in \figu{ssources} for the different discussed sources. Note that we again illustrate the size of the
relative errors as the shaded bars. Comparing \figu{ssources} to \figu{sources} illustrates that
for muon damped sources and pion beam sources the functions are decreasing in $\deltacp$ instead of increasing. Thus, a combination between $S$ and $R$ should compensate the difference from statistics between $\deltacp=0$ and $\pi$ somewhat. Comparing the error bars in these figures \figu{sources} and \figu{ssources}, there are, however, no major qualitative differences between $S$ and $R$, which means that we do not expect additional synergistic information. In particular, for neutron beams and muon damped sources, the modulation of $S$ is comparable to the one of $R$, which means that one would not expect major improvements except from increased statistics -- especially given the fact that the expected precision of $S$ may be weaker than the one of $R$ due to lower efficiencies.
The most interesting observation is probably the dependence of $S$ on $\deltacp$ for the pion beam source,
which is stronger than for $R$. Therefore, future precision measurements with high statistics (such as at IceCube volume upgrades) may find $S$ a useful observable for standard pion sources. The same applies, in principle, to the dependence on $\theta_{23}$ (not shown). We therefore focus on pion beams in this section. 

\begin{figure}[t]
\begin{center}
\includegraphics[width=8cm]{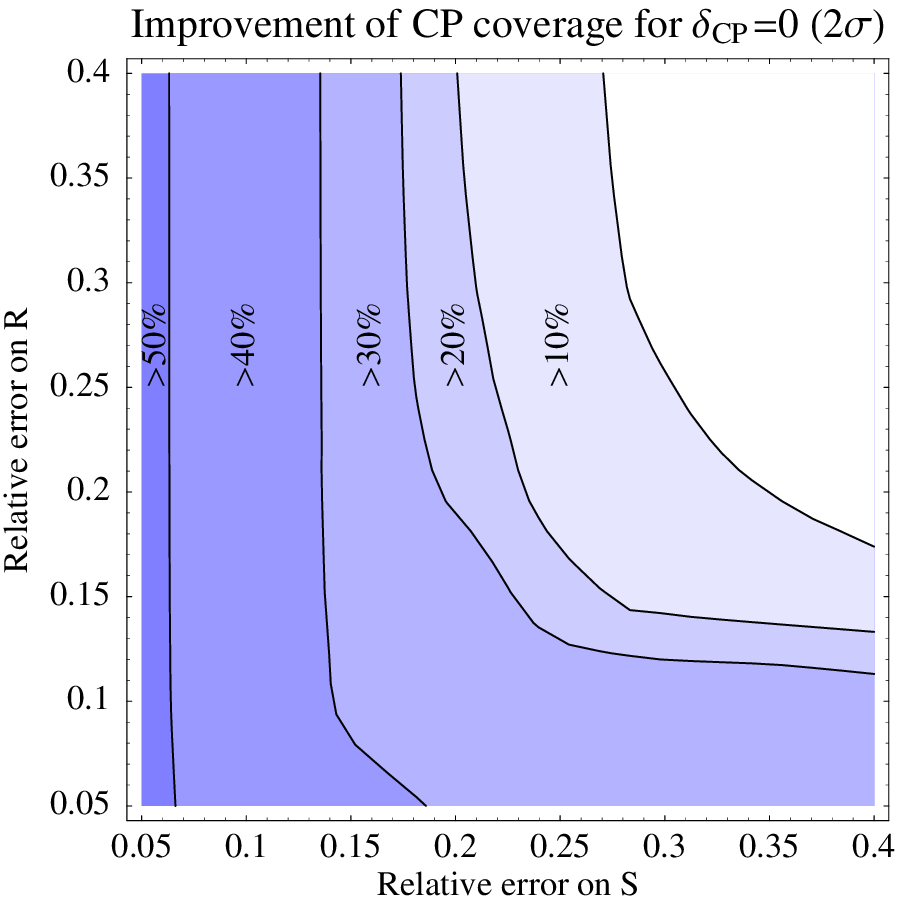}
\end{center}
\mycaption{\label{fig:cpcovsf} Relative improvement of the
CP coverage 
for $\deltacp=0$ as function of the relative errors on $S$ and $R$ for the combination of the
terrestrial experiments MINOS, Double Chooz, T2K, and NO$\nu$A with an astrophysical pion beam flux
($2 \sigma$ confidence level).}
\end{figure}

As a first example, let us study the dependence of the CP coverage on the relative errors of $S$ and $R$ from a pion beam source. Therefore,
we show in \figu{cpcovsf} the relative improvement of the CP coverage from the terrestrial 
experiments for $\deltacp=0$ at the $2 \sigma$ confidence level.  In this figure, the upper right corner (white area)
corresponds to no improvement of the terrestrial case and large errors on the astrophysical
observables, whereas the lower left corner corresponds to very precise measurements of $S$ and $R$. 
If we a assume a measurement of either $S$ or $R$, we can read off this figure the required precision
for a specific improvement. For example, for a 10\% improvement of the CP coverage, a 27\% precision 
of $S$ (intersection of 10\% contour with right vertical axis) or a 17\% precision of $R$  (intersection of 10\% contour with upper horizontal axis) are equally helpful. Obviously, the precision of $S$ has to be substantially lower (by about a factor of two) than the precision of $R$ to achieve the same result.
For a 40\% improvement, there is even a factor of three between the required precisions. One can also read off this figure that there are no real synergies between $S$ and $R$: If both quantities are measured, one cannot expect an improvement beyond what is expected from the added statistics.\footnote{For Gaussian errors, the addition of two external measurements with an equal $\Delta \chi^2$ leads to about a factor of $1/\sqrt{2}$ smaller measurement error.}

\begin{figure}[t]
\begin{center}
\includegraphics[width=8cm]{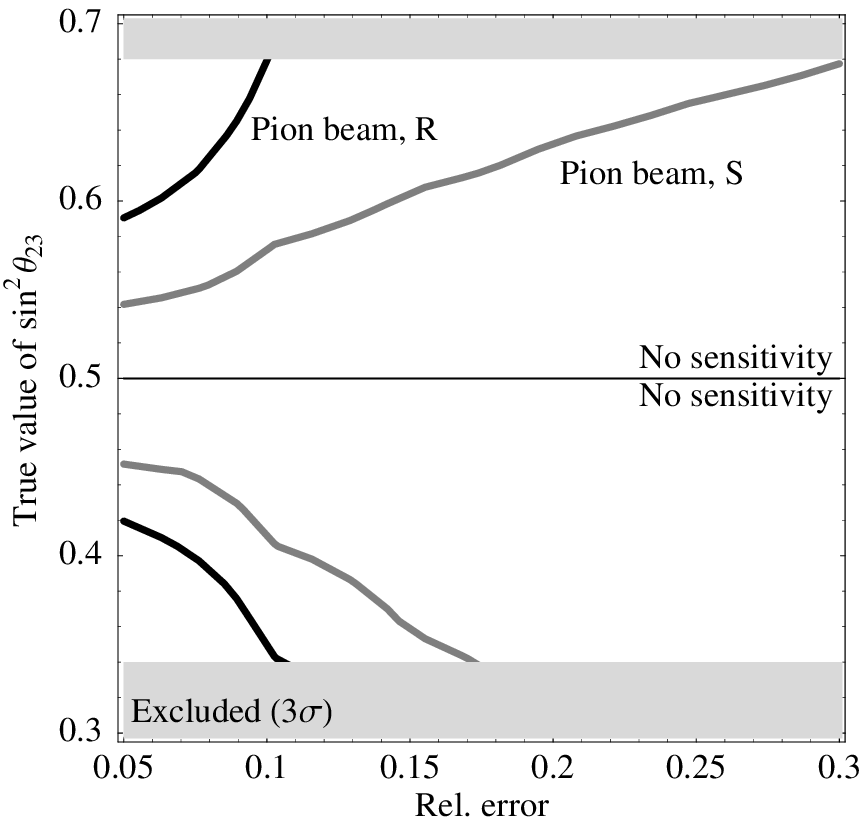}
\end{center}
\mycaption{\label{fig:theta23rs} Exclusion of the octant degeneracy as function of either the relative error on $R$ or $S$, and $\sin^2 \theta_{23}$ at the 
$2 \sigma$ confidence level for $\stheta=0$. The curves represent the terrestrial experiments (MINOS, Double Chooz, T2K, and NO$\nu$A combined) in combination
with an astrophysical pion beam flux. The gray-shaded areas 
mark the $3 \sigma$ excluded region~\cite{Maltoni:2004ei}. 
}
\end{figure}

As another example, we have indicated above that $S$ could help for the resolution of the octant degeneracy. We illustrate its impact in \figu{theta23rs}, where we show the exclusion of the octant degeneracy (same sign of $\ldm$) as function of the relative error on $S$ or $R$ at the $2 \sigma$ confidence level.
In this figure, sensitivity is given above the upper and below the lower curves. One can easily read off this figure that the precision of $S$ has to be much lower than the one of $R$ in order to do this measurement. Even a 30\% error could eliminate some this degeneracy in a part of the currently allowed $3 \sigma$ allowed region of the parameter space, and a 10\% error could eliminate the degeneracy for not too large deviations from maximal mixing.

At the end it will be a matter of statistics which of the two observables $S$ and $R$ can be measured better. In principle, if each shower in the detector (used for $R$) was identified unambiguously, the precisions of $S$ and $R$ should be approximately comparable. 
That is because the neutrinos arrive approximately in the flavor ratio (1:1:1) for this type of 
source~\cite{Learned:1994wg}, which means that comparable relative errors of $R$ and $S$ should require similar event rates
(somewhat depending on the efficiency ratio between track and shower events; see discussion in \App~\ref{app:stat}). In practice, the precisions of $S$ and $R$ may be very different because of the required higher energies for the shower identification and backgrounds from mis-identification.
Note, however, that the factor of two to three lower required precision of $S$ from physics corresponds to about a factor of four to ten lower event rate. Therefore, depending on the actual neutrino flux from the source (and the energies), $S$ could be an interesting observable as well.

\section{Summary and discussion}
\label{sec:summary}

In summary, we have studied the requirements for an astrophysical flux measurement
from a neutron beam source, muon damped source, and
pion beam source at a neutrino telescope to be relevant for the neutrino
oscillation program using terrestrial sources (reactors or accelerators). We have 
used the ratio $R= \phi_\mu/(\phi_e + \phi_\tau)$ as our main observable, which can be extracted
from the ratio between muon track and shower events in a neutrino telescope such as IceCube.
This ratio might be measured at relatively low energies at IceCube, since the threshold for muon tracks
is about $100 \, \mathrm{GeV}$, and for showers about $1 \, \mathrm{TeV}$.\footnote{However, it is
difficult to differentiate electromagnetic (from $\nu_e$) from hadronic (from $\nu_\tau$) showers
at least below several $\mathrm{PeV}$. Therefore, we have added the electromagnetic and hadronic shower events in $R$.} We have illustrated the complementarity of the observables $R$ (astrophysical fluxes), $P_{\bar{e} \bar{e}}$ (reactor experiments), and $P_{\mu e}$ or $P_{\bar{\mu} \bar{e}}$ (superbeams) with respect to their $\deltacp$-dependence: The observable of reactor experiments has no $\deltacp$-dependence, the one of first-generation superbeams depends on $\sin  \deltacp $ (at the
oscillation maximum), and the one of astrophysical neutrino sources on $\cos  \deltacp $. This complementarity has lead to a number of interesting observations for large $\stheta$: 
First, we have demonstrated that a 10\% precision
measurement from a neutron beam or muon damped source may lead to a first observation of $\deltacp$ together with Double Chooz, especially if the true $\deltacp$ is close to $\deltacp=0$ or $\pi$. This means that, depending on the time scale of the neutrino telescopes, $\deltacp$ may be restricted before the superbeams T2K and NO$\nu$A start operation. Second, we have demonstrated that a 10\% to 20\% precision of the ratio $R= \phi_\mu/(\phi_e + \phi_\tau)$ coming from a neutron beam or muon damped source, or a 5\% precision from a pion beam would be beneficial for mass hierarchy and $\deltacp$ measurements because of two factors: The $\cos \deltacp$-dependence of $R$ improves the precision of $\deltacp$ close to 
$0$ and $\pi$, and the additional restriction on the parameter space prohibits the $\mathrm{sgn}(\ldm)$-degeneracy (mainly determined by NO$\nu$A) to move in $\deltacp$-space.
For example, a moderately precise measurement of $R$ from a neutron or muon damped source
could enhance the chance to measure the mass hierarchy from about 50:50 to almost for sure.
In addition, we have found that for small values of $\stheta$, where the beams and reactor experiments
alone do not have significant sensitivity, an astrophysical neutrino flux may exclude the octant degeneracy (for large $\stheta$, reactor experiments and superbeams combined can
already resolve this degeneracy). This application may provide constraints earlier than from
atmospheric neutrinos and beams using a megaton Water Cherenkov detector~\cite{Huber:2005ep}.

Beyond the measurement of the muon track to shower ratio, we have discussed the impact of 
all-flavor identification. Therefore, we have introduced $S= \phi_e/\phi_\tau$ as an additional
observable at higher energies (possibly using detector upgrades) which could be extracted from the electromagnetic to hadronic shower ratio. We have demonstrated that this observable could be especially
interesting for pion beam sources, which might be the most likely sources to be observed
at a neutrino telescope. The required relative precision of $S$ has to be much weaker than
on $R$ for this type of source because of the stronger $\deltacp$-dependence, where the role
of $S$ is similar to the one of $R$ for  $\deltacp$, mass hierarchy, and octant degeneracy measurements.

As far as the different sources are concerned, neutron beams and muon damped sources allow for
the highest impact on the terrestrial program, which is, from the point of view of statistics,
already present for several tens of events. However, neutrinos from neutron decays have a 
flux rapidly deceasing with energy and face the atmospheric neutrino background due to the steadiness of the source~\cite{Anchordoqui:2003vc}. Muon damped fluxes, \ie, neutrinos from pion decays not followed by muon decays (which are absorbed before they can decay), occur in specific astrophysical models~\cite{Rachen:1998fd}, and pion beams may turn into these sources at high energies~\cite{Kashti:2005qa}. Therefore, neutrinos from pion decays could be the astrophysical neutrino source with most statistics. Pion beams, however, have the highest requirement with respect to precision. 
There are several arguments in favor of these sources. First, as indicated above, shower (all-flavor)
identification might help for pion beams. Second, because it may be the most likely sources,
statistics from different sources and neutrino telescopes may accumulate. And third, once such 
fluxes are detected, fiducial volume upgrades of neutrino telescopes could easily increase statistics.
Therefore, pion beams may not only have the highest requirements, but also the best prospects.

With respect to time scales, IceCube~\cite{Ahrens:2002dv} will be continuously deployed in the coming years, and other neutrino telescopes are in preparation as well~\cite{Aslanides:1999vq,Tzamarias:2003wd,Piattelli:2005hz}.
Double Chooz may have accumulated
significant data by 2011 or 2012, and the superbeams T2K and NO$\nu$A may start data taking around
that time. Beyond the coming ten years, other superbeam upgrades will follow, such as wide band beams~\cite{Borodovsky:1992pn} with a relatively wide energy spectrum. It is important to emphasize that astrophysical data can only be relevant for neutrino oscillation physics in the coming decade or at most fifteen years. 
For the discussed large values of $\stheta$, mass hierarchy, and $\deltacp$ measurements will thereafter be certainly provided by superbeam upgrades having enough statistics
(see, \eg, \Ref~\cite{Albrow:2005kw}), and the complementarity with respect to the $\cos \deltacp$-dependence will be provided by broad band superbeams, $\beta$-Beams, superbeams
operated off the oscillation maximum, or neutrino factories. For small values of $\stheta$,
atmospheric neutrinos and superbeam upgrades can exclude the octant degeneracy using megaton Water Cherenkov detectors~\cite{Huber:2005ep,Campagne:2006yx}. 

We conclude that it is worth to study the ability of
neutrino telescopes to extract the discussed observables for neutrino oscillation physics,
since this information may have a major impact on the neutrino oscillation program for the coming decade.
Of course, the exact procedure and obtainable precision may require further research and 
the actual detection of a source. In addition, the interpretation of
the data from all the discussed experiments will depend on the model of the astrophysical source.
However, important hints for the planning of future experiments may be obtained early from the
combination of a set of experiments with poor or moderate statistics each, but great synergistic potential.

\subsection*{Acknowledgments}

I would like to thank John Beacom, Carlos Pe\~na-Garay, Thomas Schwetz, and Eli Waxman
for useful discussions and comments, as well as Patrick Huber for the collaborative effort to update the NO$\nu$A simulation. In addition, I want to acknowledge support from the W.~M.~Keck Foundation and NSF grant PHY-0503584.

\begin{appendix}
\section{A note on statistics}
\label{app:stat}

\begin{table}[t]
\begin{center}
\begin{tabular}{rrrrrrr}
\hline
Muon & \multicolumn{2}{c}{Showers} & \multicolumn{4}{c}{Relative error $\Delta R/R^{th}$} \\
tracks & $\hat{\epsilon}=1$ & $\hat{\epsilon}=7$ & No BG & $\hat{\epsilon}=7$ & 10/100 BG & 100/1000 BG \\
\hline
\multicolumn{7}{l}{\underline{Neutron beam} ($R^{th} = 0.26$)} \\[0.1cm]
10 & 38 & 5 & $_{-0.33}^{+0.39}$ \\[0.1cm]
20 & 77 & 11 &  $_{-0.24}^{+0.27}$ &  $_{-0.31}^{+0.51}$ \\[0.1cm]
30 & 115 & 16 & $_{-0.20}^{+0.21}$ &  $_{-0.26}^{+0.39}$ & $_{-0.24}^{+0.29}$  \\[0.1cm]
50 & 192 & 27 & $_{-0.15}^{+0.16}$ &  $_{-0.21}^{+0.28}$ & $_{-0.18}^{+0.20}$  \\[0.1cm]
100 & 385 & 55 & $_{-0.11}^{+0.12}$ &  $_{-0.15}^{+0.19}$ & $_{-0.12}^{+0.13}$   \\[0.1cm]
250 & 962 & 137 & $_{-0.07}^{+0.07}$ &  $_{-0.10}^{+0.12}$ & $_{-0.07}^{+0.08}$ & $_{-0.13}^{+0.16}$  \\[0.1cm]
500 & 1923 & 275 & $_{-0.05}^{+0.05}$ &  $_{-0.07}^{+0.08}$ & $_{-0.05}^{+0.05}$ & $_{-0.08}^{+0.09}$  \\[0.1cm]
\hline
\multicolumn{7}{l}{\underline{Muon damped source} ($R^{th} = 0.66$)} \\[0.1cm]
10 & 15 & 2 & $_{-0.36}^{+0.50}$ \\[0.1cm]
20 & 30 & 4 &  $_{-0.26}^{+0.33}$ \\[0.1cm]
30 & 45 & 6 & $_{-0.22}^{+0.26}$ &   \\[0.1cm]
50 & 76 & 11 & $_{-0.17}^{+0.20}$ & $_{-0.26}^{+0.47}$ & $_{-0.23}^{+0.33}$  \\[0.1cm]
100 & 152 & 22 & $_{-0.12}^{+0.14}$ & $_{-0.20}^{+0.30}$ & $_{-0.15}^{+0.18}$   \\[0.1cm]
250 & 379 & 54 & $_{-0.08}^{+0.08}$ & $_{-0.13}^{+0.17}$ & $_{-0.09}^{+0.09}$ & $_{-0.23}^{+0.40}$  \\[0.1cm]
500 & 758 & 108 & $_{-0.06}^{+0.06}$  & $_{-0.10}^{+0.12}$ & $_{-0.06}^{+0.06}$ & $_{-0.13}^{+0.18}$  \\[0.1cm]
\hline
\multicolumn{7}{l}{\underline{Pion beam} ($R^{th} = 0.5$)} \\[0.1cm]
10 & 20 & 3 & $_{-0.35}^{+0.45}$  \\[0.1cm]
20 & 40 & 6 & $_{-0.25}^{+0.30}$  \\[0.1cm]
30 & 60 & 9 & $_{-0.21}^{+0.24}$ & $_{-0.30}^{+0.56}$ & $_{-0.29}^{+0.44}$  \\[0.1cm]
50 & 100 & 14 & $_{-0.16}^{+0.18}$ & $_{-0.25}^{+0.40}$ & $_{-0.21}^{+0.27}$ \\[0.1cm]
100 & 200 & 29 & $_{-0.12}^{+0.13}$ & $_{-0.18}^{+0.26}$ & $_{-0.14}^{+0.16}$  \\[0.1cm]
250 & 500 & 71 & $_{-0.08}^{+0.08}$ & $_{-0.12}^{+0.15}$ & $_{-0.08}^{+0.09}$ & $_{-0.19}^{+0.29}$ \\[0.1cm]
500 & 1000 & 143 & $_{-0.05}^{+0.06}$ & $_{-0.09}^{+0.10}$ & $_{-0.06}^{+0.06}$ & $_{-0.11}^{+0.14}$ \\[0.1cm]
\hline
\end{tabular}
\end{center}
\mycaption{\label{tab:events} Some event rates for muon tracks, the corresponding predicted shower rates
for different efficiency ratios $\hat{\epsilon}$, and relative errors  $\Delta R/R^{th}$ ($1\sigma$) for specific assumptions for backgrounds. The columns ``No BG'' and ``$\hat{\epsilon}=7$'' are calculated background-free for $\hat{\epsilon}=1$ and $\hat{\epsilon}=7$, respectively. The columns ``10/100 BG'' and ``100/1000 BG'' are calculated with backgrounds for $\hat{\epsilon}=1$, where the numbers refer to muon track background events/shower background events. In addition, a 10\% background normalization error uncorrelated between muon tracks and showers is assumed for these columns ($\sigma_b=\sigma_c=0.1$).
}
\end{table}

In this appendix, we give a simple example to illustrate the relationship between
event rates and the error on $R$. We also discuss principle challenges for the
different astrophysical neutrino sources.

We assume that all simulated rates $T_i$ and observed 
(fit) rates $O_i$ are composed of signal events $S_i$ and background events $B_i$, 
\ie,
\begin{eqnarray}
T_i & = & S_i^{th} + B_i^{th} \, , \nonumber \\
O_i & = & S^{obs}_i + B_i^{obs} \, ,
\end{eqnarray}
where $i$ corresponds to the muon tracks ($i=1$) and (non-muon track) showers ($i=2$).
Now $R^{th} \equiv \hat{\epsilon}^{-1} \, S_1^{th}/S_2^{th}$ and $R^{obs} \equiv \hat{\epsilon}^{-1} \, S^{obs}_1/S^{obs}_2$ describe the simulated and observed ratios between
muon tracks and showers. Note that these flux ratios have to be corrected by the ratio $\hat{\epsilon}$ of the efficiencies for muon track to shower detection $\hat{\epsilon} \equiv \epsilon_1/\epsilon_2$.
Usually these efficiencies should be fairly well known. They depend on the fiducial volume (which can be different for different event types), cuts to reduce backgrounds, chosen energies, \etc. For the sake of simplicity, we will assume $\hat{\epsilon}=1$ in most cases, \ie, that muon tracks and showers are identified with the same efficiencies. For IceCube, one can easily derive $\hat{\epsilon}$ from \fig~11 in \Ref~\cite{Beacom:2003nh} (see erratum) for the flux chosen therein.
One finds $\hat{\epsilon} \sim 6$ to $7$ for a muon energy threshold of $100 \, \mathrm{GeV}$, and
$\hat{\epsilon} \sim 3$ for a muon energy threshold of $1\, \mathrm{TeV}$ (muon damped or pion flux; $\hat{\epsilon} \simeq \mathrm{Muons/Showers} \cdot R^{-1}$). Therefore, we will use $\hat{\epsilon}=7$ for demonstration in some cases. Note that, in practice, the situation is more complicated and the mapping from incident to reconstructed neutrino energy needs to be described by the energy response function (usually a matrix, which describes the reconstruction at different energies). For example, some muon track events are partially contained (unless only the fully contained events are selected), which means that the muons are already loosing energy outside the detector. Thus, in general, $\hat{\epsilon}$ is a function of energy and therefore depends on the actual flux. In addition, since the muon range depends on energy (see, \eg, \fig~1 in \Ref~\cite{Beacom:2001xn}), the effective fiducial volume for partially contained events is energy dependent, as well as the event rates somewhat depend on the flavor composition. Here we assume that the energy response of the detector is known, and that $\hat{\epsilon}$ represents the integrated efficiency ratio for the flux observed.\footnote{In practice, the efficiencies for electromagnetic and
hadronic shower detection are somewhat different, too, which means that the actual observable is $R' = \phi_\mu/(\hat{\kappa} \, \phi_e + \phi_\tau)$. From \fig~11 in \Ref~\cite{Beacom:2003nh} one can extract $\hat{\kappa} \sim 0.7$ for the flux and cuts chosen therein. For this value, the relative error on $R'$ translates into the relative error on $R$ by multiplication with a factor of $\sim 1.1$ to $1.3$ for the discussed sources, whereas the qualitative dependence of $R'$ on $\stheta$ and $\deltacp$ stays the same. Since this relative efficiency depends on the instrument and analysis, we do not use it in this study.}

Let us introduce some simple systematics, too:
\begin{eqnarray}
O_1 & = & (1+a)  + (1+b) \, B_1 \, , \nonumber \\
O_2 & = & \frac{(1+a)}{\hat{\epsilon} \, R^{obs}} + (1+c) \, B_2 \, ,
\end{eqnarray}
where $a$, $b$ and $c$ are systematics auxiliary variables to be minimized
over, and $B_i=B_i^{th}$ are the predicted background events for muon tracks and showers. 
Thus, $a$ describes the unknown normalization of $S_1$ (arbitrary, therefore we have not
included it in the formula), which is,
however, fully correlated between muon track and shower events. Furthermore,
$b$ and $c$ describe the background normalization for muon tracks and showers,
respectively. Now we can define  a very simple Gaussian $\chi^2$ as 
\begin{equation}
\chi^2 = \sum\limits_{i=1}^2 \frac{(T_i-O_i)^2}{T_i} + \left( \frac{b}{\sigma_b} \right)^2 +\left( \frac{c}{\sigma_b} \right)^2 \, ,
\end{equation}
where $\sigma_b$ and $\sigma_c$ are the muon track and shower background normalization errors,
respectively, and the auxiliary variable $a$ remains unconstrained. Now we can compute the error on $R^{obs}$ for certain assumptions for the event rates $T_1$ and $T_2$ (related by $R^{th}$) and for systematics by marginalization over the systematics auxiliary variables $a$, $b$, and $c$.

As far as backgrounds are concerned, these can be very different depending on the type of the source.
Let us focus on the atmospheric neutrino background in this example. The atmospheric neutrino
background makes several ten thousands of $\nu_\mu$ events per year for $E_\nu> 100 \, \mathrm{GeV}$, 
several thousands per year for $E_\nu> 1 \, \mathrm{TeV}$, and about one hundred per year for $E_\nu> 10 \, \mathrm{TeV}$ (see, \eg, \Ref~\cite{Gonzalez-Garcia:2005xw}). Therefore, it clearly is a problem at low energies. Now there are different principle strategies to reduce this background, such as:
\begin{enumerate}
\item
 Good angular resolution
\item
 Short timescale of event
\item
 Cut low energy events
\end{enumerate}
For IceCube, the good angular resolution works for muon track events ($\sim 0.7^\circ$), but not for shower events ($\sim 25^\circ$)~\cite{Beacom:2003nh}. However, other neutrino telescopes may have a better angular resolution. The second approach helps for all short-timescale events. For instance, for gamma ray bursts, the atmospheric neutrino background should be negligible because of the short duration. 
The third approach can be useful for any source producing very high-energetic neutrinos. 
In summary, these strategies could be useful for many types of sources, such as muon damped
fluxes. However, the observation of $R$ for the neutron beams is particularly challenging because of the steadiness of the source and the flux dominating where the atmospheric neutrino background is large~\cite{Anchordoqui:2003vc}. Therefore, we conclude that the specific systematics needs further
study and clearly lies beyond the scope of this work.

We show in \Tab~\ref{tab:events} the event rates for muon tracks and showers for different hypotheses and systematics assumptions. Note that, as opposed to what is assumed for the rest of this work, the errors are often not symmetric. As one can read off this table, backgrounds become important as soon as the background rates are of the order of the signal rates. For background free measurements and $\hat{\epsilon}=1$, we find, depending on the source, a minimum of about 30 to 50 events necessary for a 20\% precision measurement, 100 to 150 events for a 10\% precision measurement, and about 500 events for a 5\% precision measurement. In this case, the statistics is determined by the muon track event rate, whereas for larger $\hat{\epsilon}$ the shower event rate becomes more important.

\end{appendix}

\end{document}